\documentclass[a4paper,11pt]{article}
\pdfoutput=1 

\usepackage{jheppub} 

\usepackage[T1]{fontenc} 

\usepackage{graphicx,verbatim,epsfig}
\usepackage{epstopdf}
\usepackage[utf8x]{inputenc}
  \usepackage{bm}
   \usepackage{amsmath}
    \usepackage{amssymb}
     \usepackage{pifont}
      \usepackage{simplewick}

\newcommand{\nn}{\nonumber}

\newcommand{\ot}{\leftarrow}

\renewcommand{\(}{\left(}
\renewcommand{\)}{\right)}
\renewcommand{\[}{\left[}
\renewcommand{\]}{\right]}

\renewcommand{\vec}[1]{\bm{#1}}

\newcommand{\bn}{{\bar n}}

\newcommand{\be}{\begin{equation}}
\newcommand{\ee}{\end{equation}}
\newcommand{\bea}{\begin{eqnarray}}
\newcommand{\eea}{\end{eqnarray}}
\newcommand{\balign}{\begin{align}}
\newcommand{\ealign}{\end{align}}

\newcommand{\sandwich}[3]{\left< #1 \right | #2 \left | #3 \right >}

\newcommand{\bg}{\begin{gather}}
\newcommand{\foma}{\end{gather}}

\newcommand{\noopsort}[1]{}

\newcommand{\vecb}[1]{\mbox{\boldmath $#1$}}

\def\<{\langle}
\def\>{\rangle}

\def\b{\beta}

\def\m{\mu}

\def\({\left(}
\def\[{\left[}
\def\){\right)}
\def\]{\right]}

\def\ln{\hbox{ln}}

\def \le { \left    }
\def \ri { \right }

\title{Power corrections and renormalons in Transverse Momentum Distributions}

\author[a]{Ignazio Scimemi}
\author[b]{and Alexey Vladimirov}

\affiliation[a]{Departamento de F\' isica Te\'orica II, Universidad Complutense de Madrid,\\
Ciudad Universitaria, 28040 Madrid, Spain}
\affiliation[b]{Institut f\"ur Theoretische Physik, Universit\"at Regensburg,\\
D-93040 Regensburg, Germany}

\emailAdd{ignazios@fis.ucm.es} \emailAdd{aleksey.vladimirov@gmail.com}

\abstract{We study  the power corrections to Transverse Momentum Distributions (TMDs) by analyzing renormalon divergences of the perturbative series. The renormalon divergences arise independently in two constituents of TMDs: the rapidity evolution kernel and the small-b matching coefficient. 
The renormalon contributions (and consequently  power corrections  and non-perturbative
corrections to the related cross sections)
have a non-trivial dependence on the Bjorken variable and the transverse distance. We discuss the consistency requirements for power corrections for TMDs and suggest inputs for the TMD phenomenology in accordance with this study. Both unpolarized quark TMD parton distribution function and fragmentation function are considered.}

\begin{document}
\maketitle
\flushbottom

\section{Introduction}

The transverse momentum dependent (TMD) distributions are fundamental non-perturbative objects that appear in many relevant processes at LHC, EIC, and $e^+e^-$ colliders, like Vector Boson Production, Higgs production, Semi-Inclusive Deep Inelastic Scattering, $e^+e^-\to 2$ hadrons. The factorization theorems which establish the definitions of TMD distributions in QCD and/or in effective field theory have been formulated recently in 
\cite{Collins:2011zzd,GarciaEchevarria:2011rb,Echevarria:2012js,Echevarria:2014rua}, using different regularization schemes. 

The perturbative properties of unpolarized TMDs, such  as evolution and operator product expansion (OPE) in the regime of small transverse momentum separation, have been  deduced  by several groups using different frameworks (see e.g.~\cite{Collins:2011zzd,GarciaEchevarria:2011rb,Chiu:2012ir,Echevarria:2014rua, Aybat:2011zv,Vladimirov:2014aja,Becher:2010tm,Ritzmann:2014mka}). The explicit direct calculation of the TMD evolution function ${\cal D}$ at NNLO has been provided in \cite{Echevarria:2015byo,Luebbert:2016itl} and recently it was obtained at N$^3$LO~\cite{Li:2016ctv}. Therefore, nowadays the perturbative knowledge of the unpolarized TMDs parton distribution functions (PDFs) and fragmentation functions (FFs) is comprehensive, thanks to the results obtained by various groups \cite{Catani:2011kr,Catani:2012qa,Catani:2013tia,Gehrmann:2012ze,Gehrmann:2014yya,Luebbert:2016itl,Echevarria:2015usa,Echevarria:2016scs}

On the contrary, the study of the non-perturbative properties of TMDs has been  based mainly on phenomenological arguments which combine the perturbative information on TMDs with their perturbatively incalculable part \cite{Aybat:2011zv,Aybat:2011ta,Sun:2013hua,Echevarria:2014rua,Boer:2014tka,Echevarria:2012pw,
Aidala:2014hva,Collins:2014jpa}. These works have lead to different forms of implementation of TMDs which in general are not easy to compare. For instance, on one hand, the well-known phenomenological considerations of Drell-Yan by ~\cite{Landry:2002ix} and~\cite{Konychev:2005iy} (the so-called BLNY model) implement an ansatz within the standard CSS approach with $b^*$-prescription in the impact parameter space (or $b$-space). They introduce a set of non-perturbative parameters $g_{1,2,3}$ and all these parameters (including the definition of $b^*$ prescription) are fundamental for these fits. The same model is also the core of the RESBOS program package \cite{RESBOS}  which is widely used in applications. On another hand, the implementation of TMDPDFs by \cite{D'Alesio:2014vja} does not use $b^*$-prescription. They have found that part of the non-perturbative corrections (essentially to the TMD evolution kernel) are negligible. They were able to describe the same data with a different shape of non-perturbative input parameterized by two parameters $\lambda_{1,2}$. Fits by other groups that limited themselves to the analysis of Vector Boson Production and Higgs production are less sensitive to the non-perturbative input (although it is still necessary) \cite{Catani:2015vma,Becher:2011xn}. Additional problems arise in the consideration of TMDFFs which are known to have very different and/or incomparable (in comparison to TMDPDFs) non-perturbative input.

This work is devoted to the study of the leading power corrections to TMD distributions.  With this aim, we perform an analysis of the leading renormalon structure of TMD distributions. A renormalon analysis of the perturbative series gives an important check of theoretical consistency for any phenomenological ansatz, although it cannot give too stringent restrictions on the fitting parameters. The study of renormalon poles allows to understand the asymptotic behavior of the perturbative series and to deduce the form of the leading non-perturbative corrections \cite{Korchemsky:1994is,Beneke:1995pq,Beneke:1997sr,Korchemsky:1997sy}.

An explicit analysis of the renormalon structure for TMDs has never been done to our best knowledge, although assumptions on its structure were used even before the actual field-theoretical definition of TMDs. We refer here, for instance, to the seminal work of \cite{Korchemsky:1994is} about the Sudakov factor in differential cross-section which is usually referred to justify a Gaussian behavior for the non-perturbative part of the TMD evolution kernel~\cite{Collins:2014jpa}. In order to describe this effect in the modern TMD framework, we recall that the definition of TMDs requires the combination of the Soft Function matrix element with the transverse momentum dependent collinear function. As we show in this work, the renormalon divergences arise in the perturbative consideration of both of these functions. These renormalon contributions have different physical meaning and should be treated independently. Firstly, the renormalon divergence of the soft factor results to a power correction within the TMD evolution kernel,  which are strictly universal for any TMD due to the universality of the soft factor itself. The leading power correction that we derive here is quadratic. The presence of these corrections has been shown in \cite{Becher:2013iya} by the analysis of the corrections to conformal anomaly. 
Secondly, the renormalon divergences naturally arise within the coefficients of the small-$b$ OPE. A study of those contributions gives access to the next twist corrections of small-$b$ matching and specifies the shape and the general scaling of TMD. 

The paper is built as the following. We provide the necessary concepts and definitions in Sec.~\ref{sec:notation}.
In Sec.~\ref{sec:TMDren} we perform the calculation of various TMD constituents (such as anomalous dimensions and coefficient functions) within the large-$\beta_0$ approximation. In the end of this section we provide a collection of the main lessons, that follows from our results. The impact of the renormalon divergences on the perturbative series and renormalon subtracted series are studied in Sec.~\ref{sec:Ren}. One of the main outcomes of the study, namely a consistent ansatz for TMDs is presented in Sec.~(\ref{sec:MODEL}).

\section{Notation and Basic Concepts}
\label{sec:notation}

Throughout the paper we follow the notation for TMDs and corresponding functions introduced in \cite{Echevarria:2016scs}. The quark TMDPDFs and TMDFFs are given by the following matrix elements
\begin{eqnarray}\nn
&&F_{q\ot N}(x,\vec b;\zeta,\mu)= \frac{Z_q(\zeta,\mu) R_q(\zeta,\mu)}{2}
\\\nn && \qquad\qquad \times
\sum_X\int \frac{d\xi^-}{2\pi}e^{-ix p^+\xi^-} \langle N|
\left\{T\[\bar q_i \,\tilde W_n^T\]_{a}\( \frac{\xi}{2} \) |X\rangle \gamma^+_{ij}\langle X|\bar T\[\tilde
W_n^{T\dagger}q_j\]_{a}\(-\frac{\xi}{2} \) \right\} | N\rangle,
\\
\label{def:TMDs}
&&\Delta_{q\rightarrow N}(z,\vec b)=\frac{Z_q(\zeta,\mu) R_q(\zeta,\mu)}{4 z N_c}
\\\nn&&
\qquad\qquad\times
\sum_X\int \frac{d\xi^-}{2\pi}e^{-ip^+\xi^-/z}\langle 0|T\[\tilde W_n^{T\dagger}q_j\]_{a}\(\frac{\xi}{2}\)
|X,N\rangle \gamma^+_{ij} \langle X,N|\bar T\[\bar q_i \,\tilde W_n^T\]_{a}\(-\frac{\xi}{2}\)|0\rangle,
\end{eqnarray}
where $R_q$ and $Z_q$ are rapidity and ultraviolet renormalization constants, $q$ are quark fields and $W^T$ are Wilson lines, and $\xi=\{0^+,\xi^-,\vec b\}$. The TMDs depend on the Bjorken variables ($x$ for TMDPDFs and $z$ for TMDFFs), the impact parameter $\vec b$ and the factorization scales $\zeta$ and $\mu$. The considerations of the TMDPDF and TMDFF are similar in many aspects. Therefore, in order to keep the description transparent we mostly concentrate on the case of the TMDPDFs, while the results for TMDFFs are presented without derivation.

The dependence on the factorization scales is given by the evolution equations, which are the  same for TMDPDF and TMDFF, namely
\begin{eqnarray}\label{def:RGE_mu_TMD}
\frac{d}{d \ln \mu^2}F_{q\ot N}(x,\vec b;\zeta,\mu)&=&\frac{\gamma^q(\mu,\zeta)}{2}F_{q\ot N}(x,\vec b;\zeta,\mu),
\\\label{def:RGE_zeta_TMD}
\frac{d}{d \ln \zeta}F_{q\ot N}(x,\vec b;\zeta,\mu)&=&-\mathcal{D}^q(\mu,\vec b)F_{q\ot N}(x,\vec b;\zeta,\mu).
\end{eqnarray}
Through the article we consider only the quark TMDs, therefore in the following we suppress the subscript $q$ on the anomalous dimensions. The values for both anomalous dimensions can be deduced from the renormalization constants \cite{Echevarria:2016scs}. Also $\gamma$ and $\mathcal{D}$ are related to each other by the cross-derivatives
\begin{eqnarray}
\frac{d \mathcal{D}(\mu,\vec{b}_T)}{d\ln \mu^2}=-\frac{1}{2}\frac{d \gamma(\mu,\zeta)}{d\ln \zeta}=\frac{\Gamma_{cusp}}{2},
\label{eq:cusp1}
\end{eqnarray}
where $\Gamma_{cusp}$ is the honored cusp anomalous dimension. 

The solution of the evolution equations Eq.~(\ref{def:RGE_mu_TMD},\ref{def:RGE_zeta_TMD}) is
\begin{eqnarray}\label{def:evol_kernel_R}
F(x,\vec b;\zeta_f,\mu_f)={\cal R}(\vec b;\zeta_f,\mu_f,\zeta_i,\mu_i)F(x,\vec b;\zeta_i,\mu_i),
\end{eqnarray}
where ${\cal R}$ is the evolution kernel,
\begin{eqnarray}
\label{eq:tmdkernelf}
\mathcal{R}(\vec b;\zeta_f,\mu_f,\zeta_i,\mu_i) &=&
\exp\bigg\{
\int_{\mu_i}^{\mu_f} \frac{d\mu}{\mu} \gamma\(\alpha_s(\mu),\ln\frac{\zeta_f}{\mu^2} \)\bigg\}
\( \frac{\zeta_f}{\zeta_i} \)^
{-\mathcal{D}(\mu_i,\vec b)}.
\end{eqnarray}
The final values of scaling parameters is dictated by the kinematic of the TMD cross-section. The variable $\zeta_f\sim Q^2$ (with $Q$ being a typical hard scale) is the scale of the rapidity factorization, and the variable $\mu_f$ is the scale of hard subprocess factorization. The intriguing point is that the evolution kernel $\mathcal{R}$ is not entirely perturbative, but contains a non-perturbative part. An estimate of the non-perturbative contribution to $\mathcal{R}$ is necessary in order to obtain the cross section in the momentum space where it is usually measured.

The non-perturbative  part of the evolution kernel is encoded in the $\mathcal{D}$-function which can be obtained from the rapidity renormalization constant $R_q$. The definition of the rapidity renormalization constant differs from scheme to scheme. In this work we use the $\delta$-regularization scheme defined in  \cite{Echevarria:2015byo,Echevarria:2016scs}. In this scheme, the $\delta$-regularization is used to regularize the rapidity divergences, and the dimensional regularization regularizes the rest of divergences. Such a configuration appears to be very effective for the TMD calculus. In particularly, the rapidity renormalization factor $R_q$ is expressed via the soft factor $S$ as $R_q=S^{-1/2}$ \cite{Echevarria:2016scs}. In the coordinate space the soft factor is given by the following matrix element
\begin{eqnarray} \label{eq:SF_def}
\tilde S(\vecb b_{T})
=
\frac{{\rm Tr}_c}{N_c} \sandwich{0}{\, T\le[S_n^{T\dagger} \tilde S_\bn^T \ri](0^+,0^-,\vecb b) \bar
T\le[\tilde S^{T\dagger}_\bn S_n^T\ri](0)}{0}
\,,
\end{eqnarray}
where we explicitly denote the ordering of operators and $S^T$ are Wilson lines, as defined in \cite{Echevarria:2015byo}. 
Considering the relation between renormalization constants one can show \cite{Echevarria:2015byo}, that 
\begin{eqnarray}\label{eq:D_fromS}
{\cal D}=\frac{1}{2}\frac{d \ln \tilde S}{d\mathbf{l}_\delta}\Bigg|_{\epsilon\rm{-finite}}
\end{eqnarray}
where $\mathbf{l}_\delta=\ln(\mu^2/|\pmb \delta |)$. Eq.~(\ref{eq:D_fromS}) can be used as the formal definition of the TMD evolution function ${\cal D}$. In this way, a non-perturbative  calculation of the SF gives access to the non-perturbative structure of  ${\cal D}$. The soft function is perturbatively universal for both Semi Inclusive Deep Inelastic Scattering and Drell-Yan type processes. Therefore, the perturbative part of the anomalous dimension $\mathcal{D}$ is universal for TMDPDF and TMDFF. One can also expect its universality in the non-perturbative regime.

The TMDs are entirely non-perturbative functions. They cannot be evaluated in perturbative QCD, due to the non-perturbative origin of hadron states. The main subject of the paper is the dependence of TMDs on the parameter $\vec b$ which is generically unrestricted since it is a variable of Fourier transformation. 
However it is interesting and numerically important to consider the range of small $b$ (here and later $b=\sqrt{\vec b^2}$). In this range, the TMDs can be matched onto corresponding integrated parton distributions. At the operator level, the small-$b$ matching is given   by the leading term of the small-$b$ OPE. The small-$b$ OPE is a formal operator relation, that relates operators with both light-like and space-like field separation to  operators with only light-like field separation. It reads
\begin{eqnarray}\label{OPE_formal}
O(\vecb b)&=&\sum_n C_n(\vecb b,\mu_b)\otimes O_n(\mu_b),
\end{eqnarray}
where $C_n$ are Wilson coefficient functions, the $\mu_b$ is the scale of small-$b$ singularities factorization or the OPE matching scale
(for simplicity we omit in Eq.~(\ref{OPE_formal}) other matching scales included in the definitions of each component of this equation). Generally, the operators $O_n$ are all possible operators with proper quantum numbers and can be organized for instance according to a power expansion, i.e. twists. In this case, the matching coefficients behave as
\begin{eqnarray}\label{coeff_powers}
C_n(\vecb b,\mu_b)\sim \(\frac{b}{B}\)^nf(\ln(\vec{b}^2 \mu^2_b)),
\end{eqnarray}
where $f$ is some function. The value of the parameter $B$ is unknown, and its origin is entirely non-perturbative. In other words, the unknown scale $B$ represents some characteristic transverse size of interactions inside a hadron $B\simeq {\cal O}(1 {\rm GeV})$. In practice it is reasonable to consider only the  leading term ($n=0$) of Eq.~(\ref{OPE_formal}) for $b\ll B$ . In this case, $f$ is an integrated parton distribution (or fragmentation function), and coefficient function is called the matching coefficient. So far, the power suppressed terms in Eq.~(\ref{OPE_formal}) has been not considered, to our best knowledge. 

For completeness, we recall here the renormalization group properties of the TMD Wilson coefficients that we use in the following sections.
The evolution equations for the matching coefficients (at $\mu_b=\mu$) with respect to $\zeta$ is
\begin{eqnarray}
\label{RGE:zeta}
\frac{d}{d\ln \zeta}C_{f\ot f'}(x,\vecb{b}_T;\mu,\zeta)=-\mathcal{D}^f(\mu,\vecb{b}_T) C_{f\ot f'}(x,\vecb{b}_T;\mu,\zeta),
\end{eqnarray}
where $f=q,g$ species, $C_{f\ot f'}$ are the matching coefficients on PDFs. It is practically convenient to extract the $\zeta$-dependence from the matching coefficient. We introduce the notation
\begin{eqnarray}\label{RGE:zeta_dep}
C_{f\ot f'}(x,\vecb{b}_T;\mu,\zeta)&=& \exp\(-\mathcal{D}^f(\mu,\vecb{b}_T)\mathbf{L}_{\sqrt{\zeta}}\)
\hat C_{f\ot f'}(x,\mathbf{L}_\mu).
\end{eqnarray}
Here and further we use the following notation for logarithms
\begin{eqnarray}\label{def:logs}
\mathbf{L}_X=\ln \(\frac{X^2 \vec b^2}{4 e^{-2\gamma_E}}\),\qquad \mathbf{l}_X=\ln \(\frac{\mu^2}{X}\).
\end{eqnarray}
The $\zeta$-free coefficient function $\hat C$ satisfies the following renormalization group equation
\begin{eqnarray}\label{RGE:C_mu}
\mu^2 \frac{d}{d\mu^2}\hat C_{f\ot f'}(x,\mathbf{L}_\mu)=\sum_r \int_x^1 \frac{dy}{y}\hat C_{f\ot r}\(\frac{x}{y},\mathbf{L}_\mu\) K_{r\ot f'}^f(y,\mathbf{L}_\mu),
\end{eqnarray}
where the kernel $K$ is
\begin{eqnarray}\nn
K^f_{r\ot f'}(x,\mathbf{L}_\mu)&=&\frac{\delta_{rf'}\delta(1- x)}{2}\(\Gamma_{cusp}^f\mathbf{L}_\mu-\gamma_V^f\)-P_{r\ot f'}(x),
\end{eqnarray}
and $P(x)$ is the splitting function (DGLAP kernel). The matching coefficient for TMDFF $\mathbb{C}_{f\to f'}$ satisfies the same set of evolution equation with only substitution of PDF splitting function $P(x)$ by the FF ones, $\mathbb{P}(z)/z^2$ \cite{Echevarria:2016scs}.
Using these equations one can find the  expression for the logarithmic part of the matching coefficients at any given order, in terms of the anomalous dimensions and the finite part of the coefficient functions. The expressions for the anomalous dimensions, the recursive solution of the RGEs and the explicit expressions for the coefficients $C$ and $\mathbb{C}$ can be found, e.g. in \cite{Echevarria:2016scs}. 

\section{TMD in large-$\beta_0$ approximation and renormalon divergences}
\label{sec:TMDren}

The leading non-perturbative contribution to the perturbative series is commonly associated with renormalons. The renormalon contributions were intensively studied for various matrix elements and in different regimes, for review see \cite{Beneke:1998ui,Beneke:2000kc}. A typical signature of renormalons is the factorial divergence of the perturbative series. These divergences are often discussed in terms of the corresponding singularities in the Borel plane. 

The best representative and the only stable way to study the renormalon divergence within perturbative QCD is the large-$\beta_0$ approximation. The large-$\beta_0$ expression can be obtained from the large-$N_f$ expression through the procedure of "naive Abelianization" \cite{Ball:1995ni,Beneke:1994qe}.  In this section, we present the calculation of large-$\beta_0$ correction to TMDs. Since the technique of large-$N_f$ calculus is well-known, we skip the detailed evaluation (redirecting the reader to the related literature) and present only intermediate expressions.

\subsection{The soft function in the large-$\beta_0$ approximation}

The soft function matrix  elements is a key structure  for the TMD construction and as such it is a  good starting point for the renormalon analysis. The large-$\beta_0$ calculation of the soft factor runs in  parallel to the calculation of the integrated soft factor for Drell-Yan, which is presented in \cite{Beneke:1995pq} (see Sec.5.3). Here we present our results of the evaluation.

\begin{figure}[t]
\centering
\includegraphics[width=0.55\textwidth]{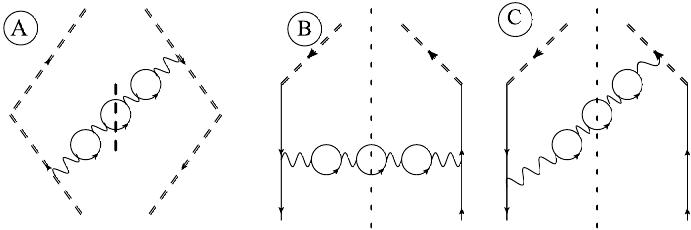}
\caption{Diagrams contributing to the leading order of large-$N_f$ limit. The diagram $A$ is the contribution to the soft factor. Diagrams $B$
and $C$ are contribution to the matching coefficient. The counter term diagrams are not shown.} \label{fig:1loop}
\end{figure}

To begin with, we evaluate the large-$N_f$ contribution to the soft factor, which is given by the "bubble" resummed diagram, shown in Fig.\ref{fig:1loop}.A. The expression for the (renormalized) diagram with $n$-bubble insertion  is
\begin{eqnarray}\label{SF:at_large_N}
\text{SF}_n&=&-\frac{4C_F}{\beta_0^f}\(\frac{a_s\beta_0^f}{-\epsilon}\)^{n+1}\sum_{k=0}^{n}\frac{n!}{k!(n-k)!}
\\\nn&& \frac{(-1)^{k}}{n-k+1}G(-\epsilon,-(n+1-k)\epsilon)\(
\mathbf{L}_{\bm\delta}-\psi(-(n-k+1)\epsilon)-\gamma_E\),
\end{eqnarray}
where $\beta_0^f=\frac{4}{3}T_rN_f$, $a_s=g^2/(4\pi)^2$, $\epsilon$ is the parameter of dimension regularization ($d=4-2\epsilon$), $\pmb \delta=|2\delta^+\delta^-| $ with $\delta^{+(-)}$ the being parameters of rapidity regularization for Wilson lines pointing in $n$($\bar n$)-direction \cite{Echevarria:2016scs}. The function $G$ is a standard function that appears in the large-$\beta_0$ calculation \cite{Ball:1995ni,Beneke:1994qe,Beneke:1998ui,Beneke:1995pq}, and  is given by the expression 
\begin{eqnarray}
G(\epsilon,s)=e^{s\gamma_E}\pmb B_\mu^{-s} A_{-\epsilon}^{s/\epsilon-1}\frac{\Gamma(1+s)}{\Gamma(1-s+\epsilon)},
\end{eqnarray}
with 
$$
A_\epsilon=\frac{6\Gamma(1+\epsilon)\Gamma^2(2-\epsilon)}{\Gamma(4-2\epsilon)},\qquad \pmb B_\mu=\frac{\vec b^2\mu^2}{4 e^{-2\gamma_E}}.
$$
Here, the Euler-Mascheroni constant is a result of the $\overline{\text{MS}}$ scheme. For
$n=1,2$ this expression  agrees with the direct calculation of the soft factor in $\delta$-regularization \cite{Echevarria:2015byo}. We also introduce an additional function for the double-pole part
\begin{eqnarray}
\tilde G(\epsilon,s)=-G(\epsilon,s)(\psi(s)+\gamma_E).
\end{eqnarray}
The functions $G$ and $\tilde G$ have the following Taylor series
\begin{eqnarray}
G(\epsilon,s)&=&\sum_{j=0}^\infty G_j(\epsilon)s^j=\sum_{j=0}^\infty s^j \sum_{k=0}^\infty g_k^{[j]}\epsilon^k,
\\
\tilde G(\epsilon,s)&=&\sum_{j=0}^\infty \tilde G_j(\epsilon)s^{j-1}=\sum_{j=0}^\infty s^{j-1} \sum_{k=0}^\infty \tilde g_k^{[j]}\epsilon^k.
\end{eqnarray}
These expressions define the coefficients $g_k^{[j]}$ and $G_j$. Note, that $g_k^{[0]}= \tilde g_k^{[0]}$ and $g_k^{[1]}= \tilde
g_k^{[1]}$.

The procedure of "naive Abelianization" consists in the replacement of $N_f$ by the corresponding $\beta_0$ expression \cite{Beneke:1994qe}, i.e.
\begin{eqnarray}\label{naive_abel}
\beta_0^f=\frac{4}{3}T_rN_f\qquad \longrightarrow \qquad -\beta_0=-\frac{11}{3}C_A+\frac{4}{3}T_rN_f.
\end{eqnarray}
In this way, we obtain the large-$\beta_0$ expression for the soft factor
\begin{eqnarray}\label{SF:result}
\text{SF}&=&-\sum_{n=0}^\infty \frac{4C_Fc_s^{n+1}}{\beta_0} \Bigg[(-1)^n n!\(\mathbf{L}_{\pmb \delta} G_{n+1}(-\epsilon)+\tilde G_{n+2}(-\epsilon)\)
\\\nn &&
+\frac{(-1)^n}{n+1}\(-\mathbf{L}_{\pmb \delta}\frac{G_0(-\epsilon)}{\epsilon^{n+1}}+\frac{\tilde G_0(-\epsilon)}{\epsilon^{n+2}}(\psi(n+2)+\gamma_E)
-\frac{\tilde G_1(-\epsilon)}{\epsilon^{n+1}}\)\Bigg],
\end{eqnarray}
where we have introduced the  large-$\beta_0$ coupling constant
$$c_s=\beta_0 a_s>0.$$
Note, that in Eq.~(\ref{SF:result}) the terms suppressed in $\epsilon$ are dropped. 

Eq.~(\ref{SF:result}) gives access to the anomalous dimension $\mathcal{D}$, which we study in Sec.~\ref{sec:renorm_for_AD}, and to the rapidity renormalization factor $R_q$. The factor $R_q$ (we recall that it is equal to $R_q=S^{-1/2}$ in the $\delta$-regularization \cite{Echevarria:2016scs}) from the perspective of the large-$\beta_0$ approximation has the same perturbative combinatorics as the one-loop-truncated pertrubation series. It is given by
\begin{eqnarray}\label{def:R_at_largeb}
R_q=1-\frac{\text{SF}}{2}
\end{eqnarray}
at $\delta^-=\zeta/p^+$ and $\text{SF}$ given in Eq.~(\ref{SF:result}). This expression is used in the next section to extract the large-$\beta_0$ expression of the Wilson coefficients of small-$b$ OPE.

\subsection{The TMD  in the large-$\beta_0$ approximation}

To obtain the TMD matching coefficient one should evaluate the diagrams B and C, which are shown in 
Fig.~\ref{fig:1loop}. The
result for the sum of these  diagrams and their Hermitian conjugations is
\begin{eqnarray}
\label{eq:phi}
\Phi_{q\ot q}&=&\frac{2C_F}{\beta_0^f}\sum_{n=0}^\infty
\frac{(a_s\beta_0^f)^{n+1}}{(-\epsilon)^{n+1}}\sum_{k=0}^{n}\frac{n!}{k!(n-k)!}\frac{(-1)^{k}G(-\epsilon,-(n-k+1)\epsilon)}{n-k+1}
\\\nn&&\qquad\qquad
\Bigg[\bar x x^{(n-k)\epsilon}(1-\epsilon)(1+(n-k)\epsilon)+2\frac{x^{1+(n-k)\epsilon}}{(1-x)_+}-2\delta(\bar
x)\ln\(\frac{\delta^+}{p^+}\)\Bigg],
\end{eqnarray}
where we have used the same notation as in Eq.~(\ref{SF:at_large_N}) and $\bar x=1-x$. The last term in square brackets represents the rapidity divergence which appears in the diagram $C$. For $n=0,1$ this expression reproduces the result of explicit calculation made in \cite{Echevarria:2015usa}.

Using  Eq.~(\ref{def:R_at_largeb}) and Eq.~(\ref{eq:phi})  we can complete the result for the large-$N_f$ expression of the TMDPDF,
\begin{eqnarray}\label{TMDPDF:largeN1}
R_q\Phi&=&\Phi_{q\ot q}-\frac{\text{SF}}{2}=\frac{2C_F}{\beta_0^f}\sum_{n=0}^\infty
\frac{(a_s\beta_0^f)^{n+1}}{(-\epsilon)^{n+1}}\sum_{k=0}^{n}\frac{n!}{k!(n-k)!}\frac{(-1)^{k}G(-\epsilon,-(n-k+1)\epsilon)}{n-k+1}
\\\nn&&
\Bigg[\bar x x^{(n-k)\epsilon}(1-\epsilon)(1+(n-k)\epsilon)+2\frac{x^{1+(n-k)\epsilon}}{(1-x)_+}+\delta(\bar
x)\(\mathbf{L}_\mu-\mathbf{l}_\zeta-\psi(-(n-k+1)\epsilon)-\gamma_E\)\Bigg].
\end{eqnarray}
Here, we observe the cancellation of the rapidity divergences that leaves the residual $\mathbf{l}_\zeta$ dependence.

In order to extract the  matching coefficient of the TMDPDF onto the PDF one has to proceed to the renormalization of Eq.~(\ref{TMDPDF:largeN1}). This is greatly simplified  in the $\delta$-regularization scheme, where all virtual graphs and integrated graphs are zero. The only non-zero contribution is the UV counterterm which is a pure $\epsilon$-singularity. The accounting of this part eliminates terms singular in $\epsilon$, leaving the finite part unchanged. The latter provides the coefficient function. Performing the "naive Abelianization" as in Eq.~ (\ref{naive_abel}) we obtain the large-$\beta_0$ result
\begin{eqnarray}\label{TMDPDF:largeb}
C_{q\ot q}&=&\frac{2C_F}{\beta_0}\sum_{n=0}^\infty c_s^{n+1}\Bigg\{
\Bigg[\bar x +2\frac{x}{(1-x)_+}\Bigg]\Bigg[\frac{\gamma_{n+1}(x)}{n+1}+(-1)^nn!g^{[n+1]}_0[ {\pmb B}_{\sqrt{x}\mu}]\Bigg]
\\\nn&&
+\frac{\bar x}{n+1}\(2\gamma_n(x)+\gamma_{n-1}(x)\)-\bar x(-1)^nn!g^{[n]}_0[ {\pmb B}_{\sqrt{x}\mu}]
\\\nn&& +\delta(\bar x)\(\mathbf{L}_\mu-\mathbf{l}_\zeta\)\Bigg[\frac{g^{[0]}_{n+1}}{n+1}+(-1)^nn!g^{[n+1]}_0\Bigg]
\\\nn
&&+\delta(\bar x)\Bigg[\tilde g^{[0]}_{n+2}\frac{\psi(n+2)+\gamma_E}{n+1}+\frac{\tilde g^{[1]}_{n+1}}{(n+1)}+(-1)^nn!\tilde g^{[n+2]}_0 \Bigg] \Bigg\},
\end{eqnarray}
where $\pmb B_{\sqrt{x}\mu}=x\pmb B_\mu$, and
$$
x^\epsilon G_0(\epsilon)=\sum_{k=0}^\infty \gamma_k\epsilon^k.
$$
The additional variable in the square brackets for the functions $g$ indicates the modified value of $\pmb B_\mu$ to be substituted.

The calculation of TMDFFs matching coefficient proceeds in the same  way as for TMDPDFs. The result of the calculation is
\begin{eqnarray}\label{TMDFF:largeb}
z^2\mathbb{C}_{q\ot q}&=&\frac{2C_F}{\beta_0}\sum_{n=0}^\infty c_s^{n+1}\Bigg\{
\Bigg[\bar z +2\frac{z}{(1-z)_+}\Bigg]\Bigg[\frac{\gamma_{n+1}(z^{-1})}{n+1}+(-1)^nn!g^{[n+1]}_0[ {\pmb B}_{\mu/\sqrt{z}}]\Bigg]
\\\nn&&
+\frac{\bar z}{n+1}\(2\gamma_n(z^{-1})+\gamma_{n-1}(z^{-1})\)-\bar z(-1)^nn!g^{[n]}_0[ {\pmb B}_{\mu/\sqrt{z}}]
\\\nn&& +\delta(\bar z)\(\mathbf{L}_\mu-\mathbf{l}_\zeta\)\Bigg[\frac{g^{[0]}_{n+1}}{n+1}+(-1)^nn!g^{[n+1]}_0\Bigg]
\\\nn
&&+\delta(\bar z)\Bigg[\tilde g^{[0]}_{n+2}\frac{\psi(n+2)+\gamma_E}{n+1}+\frac{\tilde g^{[1]}_{n+1}}{(n+1)}+(-1)^nn!\tilde g^{[n+2]}_0 \Bigg]
\\&& \nn
-\sum_{r=1}^{n+1}\(\(\bar z+\frac{2z}{1-z}\)\gamma_{n-r+1}(z)+\bar
z(2\gamma_{n-r}(z)+\gamma_{n-r-1}(z)\)\frac{(-1)^r\ln^r(z^{2})}{(n+1)r!} \Bigg\}.
\end{eqnarray}
One can see that the expression for TMDPDF Eq.~(\ref{TMDPDF:largeb}) is related to the first four lines of the expression for TMDFF
Eq.~(\ref{TMDFF:largeb}) by the crossing relation $x\to z^{-1}$. The last line of Eq.~(\ref{TMDFF:largeb}) is specific for TMDFF and it is an effect of the expansion of the normalization factor $z^{-2\epsilon}$.

One can check that at $n=0,1$ the expressions (\ref{TMDPDF:largeb}) and (\ref{TMDFF:largeb}) coincide with the one calculated in \cite{Echevarria:2016scs}.

\subsection{The TMD anomalous dimensions at large-$\beta_0$ and renormalon singularities of $\mathcal{D}$}
\label{sec:renorm_for_AD}

In the articles \cite{Echevarria:2015byo,Echevarria:2016scs} it was shown that in the $\delta$-regularization scheme the anomalous dimension $\mathcal{D}$ can be obtained from the rapidity singular part of the soft factor as in (\ref{eq:D_fromS}). Considering the Eq.~(\ref{SF:result}) we obtain the anomalous dimension $\mathcal{D}$ in the large-$\beta_0$ approximation
\begin{eqnarray}\label{D_AD:n-expression}
\mathcal{D}&=&-\frac{2C_F}{\beta_0}\sum_{n=0}^\infty c_s^{n+1} \((-1)^n n! g^{[n+1]}_0+\frac{g_{n+1}^{[0]}}{n+1}\).
\end{eqnarray}
The first term in the brackets of Eq.~(\ref{D_AD:n-expression}) behaves $\sim n!$ at large $n$, and represents the renormalon singularity. 

At this point it is convenient to consider the Borel transformation of the result. We define the Borel transformation  of a perturbative series in the usual way
\begin{eqnarray}
f(c_s)=\sum_{n=0}^\infty f_n c_s^{n+1}\qquad &\Longrightarrow& \qquad B[f](u)=\sum_{n=0}^\infty f_n\frac{u^n}{n!}.
\end{eqnarray}
A perturbative series is Borel summable if an integral
\begin{eqnarray}
\label{eq:antiBorel}
\tilde{f}&=&\int_0^\infty du e^{-u/c_s}B[f](u),
\end{eqnarray}
exists.
Performing  the Borel transformation on the  $\mathcal{D}$ function and applying Eq.~(\ref{eq:antiBorel}), we find
\begin{eqnarray}\label{D_AD:summed}
\mathcal{D}&=&-\frac{2C_F}{\beta_0}\(\int_0^{c_s} dx\frac{G(x,0)-1}{x}-\int_0^\infty du \frac{G(0,-u)-1}{u}e^{-u/c_s}\).
\end{eqnarray}
The first term is analytical and reproduces the cusp-anomalous dimension at large-$\beta_0$ \cite{Beneke:1995pq}
\begin{eqnarray}
\Gamma_{cusp}(c_s)=\frac{4C_F c_s}{\beta_0}\frac{\Gamma(4+2c_s)}{6\Gamma^2(2+c_s)\Gamma(1+c_s)\Gamma(1-c_s)}=\frac{4C_F c_s}{\beta_0}G(c_s,0).
\end{eqnarray}
The function which appears in the second term
\begin{eqnarray}
G(0,-u)=\pmb  B_\mu^u e^{(\frac{5}{3}-2\gamma_E)u} \frac{\Gamma(1-u)}{\Gamma(1+u)},
\end{eqnarray}
contains a series of poles at $u=1,2,...$ which correspond to infrared renormalons. One can check explicitly that the relation Eq.~(\ref{eq:cusp1}) holds for large-$\beta_0$ expression, due to cancellation of the renormalon divergences in the second term of Eq.~(\ref{D_AD:summed}) between derivative of coupling constant (in the Borel exponent) and derivative of $\pmb B_\mu$ (in the function $G(0,-u)$).

There are multiple possibilities to define the sum Eq.~(\ref{D_AD:n-expression}), e.g. one can slightly shift the integration contour for Eq.~(\ref{D_AD:summed}) into the complex plane. The difference between integrals passing from the lower and upper sides of poles is called infrared (IR)-ambiguity and is given by a $(-\pi)$ times the residue at the pole. For the anomalous dimension $\mathcal{D}$ it reads
\begin{eqnarray}\label{IRunamb:D}
\delta_{IR}\{\mathcal{D}\}=c \vec b^2 \Lambda^2,
\end{eqnarray}
where
\begin{eqnarray}\label{def:c}
c=\frac{\pi C_F}{2\beta_0}e^{\frac{5}{3}}\simeq 1.2.
\end{eqnarray}
The IR-ambiguity represents the typical scale of the error for perturbative series.

The same conclusion, namely the presence  of a $\vec b^2$-correction for $\mathcal{D}$, was made in Ref.~\cite{Becher:2013iya} using different argumentation. In Ref.~\cite{Becher:2013iya} the factorized cross-section has been considered within the soft collinear effective field theory (SCET). It has been shown that the power correction to the soft factor which arises in the next-to-leading term of large-$Q^2$ OPE, is proportional to the soft factor matrix element. Exponentiating the power correction one obtains the same result as presented here. It is an expected agreement because the renormalon calculation is equivalent to the calculation of the correction term of OPE.

The anomalous dimension $\gamma_V$ can be extracted from the coefficient function Eq.~(\ref{TMDPDF:largeb}). We consider the derivative of
coefficient function at $\mathbf{l}_\zeta=\mathbf{L}_\mu$
\begin{eqnarray}\label{RG_for_coeff}
\mu^2 \frac{d}{d\mu^2}\hat C_{q\ot q}(x,\mathbf{L}_\mu)=\int_x^1 \frac{dy}{y}\hat C_{q\to q}\(\frac{x}{y},\mathbf{L}_\mu\) \(
\frac{1}{2}\(\Gamma_{cusp}\mathbf{L}_\mu-\gamma_V\)\delta(\bar y)-P_{q\ot q}(y)\),
\end{eqnarray}
where we have dropped the mixing among flavors. The DGLAP kernel at large-$\beta_0$ is given by the expression
\begin{eqnarray}
P_{q\ot q}(x)&=&\frac{2C_F}{\beta_0}\sum_{n=0}^\infty c_s^{n+1} \Bigg\{ \frac{1+x^2}{1-x}\gamma_n(x)+\bar
x(2\gamma_{n-1}(x)+\gamma_{n-2}(x))\Bigg\}_+.
\end{eqnarray}
Considering the derivative of Eq.~(\ref{TMDPDF:largeb}) and comparing right and left hand sides of Eq.~(\ref{RG_for_coeff}) we obtain
\begin{eqnarray}
\gamma_V&=&-\frac{4C_F}{\beta_0}\Bigg\{c_s\(\psi(1+c_s)+2\gamma_E+\frac{3-c_s^2}{(1+c_s)(2+c_s)}\)G(c_s,0)
\\\nn&& \qquad +\ln\(G(c_s,0)\Gamma(1+c_s)\)G(c_s,0)+\int_0^1 \frac{G(xc_s,0)-G(c_s,0)}{1-x}dx\Bigg\}.
\end{eqnarray}
This expression contains no singularity, and hence it is renormalon-free, as it is usually expected for an ultraviolet anomalous dimension.

\subsection{TMD matching coefficient at large-$\beta_0$}
\label{sec:matching_at_b0}

Before  the evaluation of the sums in Eq.~(\ref{TMDPDF:largeb}-\ref{TMDFF:largeb}) we extract the part related to the anomalous dimension $\mathcal{D}$ to obtain the coefficients $\hat C$ defined in Eq.~(\ref{RGE:zeta_dep}). This procedure is important since the function $\mathcal{D}$ contains its own renormalon singularities, as described in Eq.~(\ref{IRunamb:D}). The contribution of $\mathcal{D}$ is easily recognized in the third lines of (\ref{TMDPDF:largeb}-\ref{TMDFF:largeb}) (compare with Eq.~(\ref{D_AD:n-expression})).

The result of the Borel transform for the coefficient $\hat C$, Eq.~(\ref{TMDPDF:largeb}) is
\begin{eqnarray}\nn
\hat C&=&\frac{2C_F}{\beta_0}\int_0^\infty du e^{-u/c_s}\Bigg\{
\(\bar x+\frac{2x}{(1-x)_+}\)\frac{\pmb \gamma(u)-1}{u}+
\bar x \int_0^1 dy(2+u\bar y)\pmb \gamma(yu)
\\\nn &&+\delta(\bar x)\(\frac{\pmb G_1(u)-1}{u}+\int_0^1 dy \frac{\pmb G_0'(u)-\pmb G_0'(yu)}{u(1-y)}\)
\\\nn&&+\(\bar x+\frac{2x}{(1-x)_+}\)\(\frac{1-G[\pmb B_{\sqrt{x}\mu}](0,-u)}{u}\)-\bar x G[\pmb B_{\sqrt{x}\mu}](0,-u)
\\ &&+\delta(\bar x)
\(\frac{G(0,-u)(\psi(-u)+\gamma_E)}{u}-\frac{1}{u^2}-\frac{\pmb L_\mu+\frac{5}{3}}{u}\) \Bigg\},
\label{TMDPDF:largeb_sum}
\end{eqnarray}
where by bold font we denote the Borel transformed functions,
\begin{eqnarray}
\bm{G}_i(u)=\sum_{n=0}^\infty g^{[i]}_n\frac{u^n}{n!},~~~\pmb G_0'(x)=\frac{d}{dx}\pmb G_0(x),~~~\bm \gamma_0(u)=\sum_{n=0}^\infty
\gamma_n(x)\frac{u^n}{n!}.
\end{eqnarray}
The terms in Eq.~(\ref{TMDPDF:largeb_sum}) are collected such that every bracket is finite at $u\to 0$. The expression for TMDFF coefficient function $\hat{\mathbb{C}}$ can be obtained using the crossing transformation ($x\to z^{-1}$) and the addition of the normalization contribution (the last line in Eq.~(\ref{TMDFF:largeb})). 

In the last two lines of Eq.~(\ref{TMDPDF:largeb_sum}) we have the infrared renormalon poles in $u=1,2,..$. One can see that the
third line contains only first order poles, while the last line contains second order poles at $G(0,-u)\psi(-u)$. Considering the infrared
ambiguity at $u=1$ we obtain
\begin{eqnarray}\label{IRunamb:C}
\delta_{IR}\{\hat C\}=-c(x \vec b^2 \Lambda^2)\Bigg\{2\bar x+\frac{2x}{(1-x)_+}-\delta(\bar x)\(\mathbf{L}_\Lambda+\frac{2}{3}\) \Bigg\},
\end{eqnarray}
where constant $c$ is given in Eq.~(\ref{def:c}). The $x-$dependence of this expression \textit{exactly} reproduces the $x-$dependence of the leading terms of the next power correction in small-$b$ OPE, see detailed description in \cite{Braun:2004bu}. The consideration of ambiguites of higher renormalon poles gives access to the higher-power corrections. We obtain
\begin{eqnarray}
\label{IRunamb:Cn}
\delta_{IR}^{u=n}\{\hat C\}=\frac{\pi C_F}{\beta_0}\frac{\(-x \vec b^2 \Lambda^2 e^\frac{5}{3}\)^n}{n!n!}\(\frac{2x}{(1-x)_+}+(n+1) \bar x-\delta(\bar x)\(\mathbf{L}_\Lambda-\psi_{n+1}-\gamma_E+\frac{5}{3}\)\).
\end{eqnarray}
However, these expressions can be modified by the infrared renormalon contributions of the higher-twist terms. The most important information of the higher-power corrections is that the renormalons scale as $x\vec b^2$, but not as $\vec b^2$ which is a naive assumption. The consequences of this fact are discussed in the next sessions.

The corresponding calculation for TMDFF gives
\begin{eqnarray}\label{IRunamb:CC}
\delta_{IR}\{z^2\hat {\mathbb{C}}\}=-c\(\frac{\vec b^2 \Lambda^2}{z}\)\Bigg\{2\bar z+\frac{2z}{(1-z)_+}-\delta(\bar z)\(\mathbf{L}_\Lambda+\frac{2}{3}\)\Bigg\}.
\end{eqnarray}
which is the same as Eq.~(\ref{IRunamb:C}) with the crossing change $x\to 1/z$. One can see that the difference in normalization which spoils the crossing between TMDPDFs and TMDFFs, disappears in the renormalon contribution. The higher poles ambiguites are provided using the crossing relation  $x\to 1/z$ in Eq.~(\ref{IRunamb:Cn}).

\subsection{Lessons from large-$\beta_0$}
\label{subsec:lessons}

The  Eq.~(\ref{IRunamb:D}, ~\ref{IRunamb:C},~\ref{IRunamb:CC}) are one of the main results of this work. These expressions represent the leading power correction to the small-$b$ regime, where all perturbative properties of TMDs are derived. These expressions  give access to a general structure of the next-to-small-$b$ regime. The practical implementation of results Eq.~(\ref{IRunamb:D},~\ref{IRunamb:C},~\ref{IRunamb:CC}) is given in the next section, while here we collect the most important observation that follows from the large-$\beta_0$ calculation and which should be taken into account for TMD phenomenology.

The first, and the most obvious, observation is that the leading power corrections are $\sim \vec b^2$.  It implies that an exponential decay of the TMDs that is sometimes suggested in phenomenological studies (e.g. \cite{Schweitzer:2012hh,Collins:2013zsa}) can in no way affect the small-$b$ region. Indeed, it would imply the corrections $\sim \sqrt{\vec b^2}$ to the small-$b$ OPE, that cannot appear without extra scaling parameter. Nonetheless, exponential corrections can occur  in the large-$b$ regime, which is inaccessible by perturbative considerations. 

Second, one can see that the renormalon corrections to TMDPDFs matching coefficient scales like $x\vec b^2$, and not as simply $\vec b^2$
(as it is usually assumed), nor as $x^2\vec b^2$ (as suggested by Laguerre polynomial decomposition \cite{Vladimirov:2014aja}). Therefore, the contributions of higher-twist terms in small-$b$ OPE for TMDPDF are largely functions of $x\vec b^2$. Correspondingly, TMDFFs matching coefficients are a function of $\vec b^2/z$. This is important in respect of the phenomenological implementation of the TMDs. For instance, the $b^*$-prescription which is often adopted  does not respect  this scaling and so, in this sense, it is not  fully consistent with the estimated higher twist effects.

Third, the renormalon contributions to the anomalous dimension $\mathcal{D}$ and to matching coefficients have different physical origins and do not mix with each other.  In fact, the anomalous dimension $\mathcal{D}$ is an universal object that is the same for all regimes of $b$ and for TMDs of different quantum numbers~\cite{Collins:2014jpa}. Thus, the renormalon contribution to $\mathcal{D}$ represents a generic universal non-perturbative contribution, alike in the case of heavy quark masses.
On the other hand, the (infrared) renormalon divergences within the matching coefficients are to be canceled  by the corresponding (ultraviolet) renormalon contributions of higher twists. Therefore, while Eq.~(\ref{IRunamb:D}) represents a size of a universal non-perturbative contribution,  Eq.~(\ref{IRunamb:C},~\ref{IRunamb:CC}) give the form of the twist-four contribution to small-$b$ OPE. In other words,  Eq.~(\ref{IRunamb:C},~\ref{IRunamb:CC})  estimate very accurately the $x$-behavior of subleading correction to small-$b$ OPE.

The consideration of the anomalous dimension $\mathcal{D}$  for gluon distributions is identical to those of quarks (apart of trivial replacing of common the factor $C_F$ by $C_A$). Contrary, the calculation of the renormalon contribution for gluon and quark-gluon matching coefficient is much more complicated than the one presented here and is beyond the scope of this paper. In general, we can expect a non-trivial dependence of the renormalon contribution on the Bjorken variables.  At present, we cannot find arguments  which suggest a location  for the renormalon poles  and  an $x\vec b^2$ scaling different  from that  of quarks.

\section{Renormalon substraction and power corrections}
\label{sec:Ren}

Our analysis is limited to the  quark TMDs only. Nonetheless, we can advance some considerations on possible inputs, which are  consistent with our  findings and evaluate their impact on the non-perturbative structure of TMDs. The suggested ansatz for TMDs does not pretend to be unique and moreover is inspired by other popular models. We postpone to a future publication a more dedicated study on the subject. 

We recall here the form of the TMDPDFs which emerges at small-$b$ is
\begin{eqnarray}\label{def:pert_TMD}
&&F^{pert}_{q\ot N}(x,\vec b;\zeta_f,\mu_f)=
 {\cal R}(\vec b,\zeta_f,\mu_f;\zeta_b,\mu)\sum_j \int_x^1 \frac{dy}{y}\hat C_{q\ot j}\(\frac{x}{y},\vec
b;\mu\)f_{j\ot N}(y,\mu),
\end{eqnarray}
where  the  evolution kernel ${\cal R}$ is given in Eq.~(\ref{eq:tmdkernelf}). The argument $\zeta_b$ of ${\cal R}$ is collected from the combination of two exponents: the original factor $\mathcal{R}$ (\ref{eq:tmdkernelf}) and the exponential prefactor of $\hat C$ (\ref{RGE:zeta_dep}), and it takes the  value
$$
\zeta_b=\frac{4 e^{-2\gamma_E}}{\vec b^2}.
$$
The analogue equation for TMDFFs is obtained replacing  consistently the PDF $f_{j\ot N}$ by the fragmentation function $d_{j\to N}$ and the coefficient function $C_{q\ot j}$ by $\mathbb{C}_{q\to j}$, while the evolution kernel remains the same. This expression is usually taken as an initial ansatz for TMD phenomenology.

As we pointed earlier there are two places where the non-perturbative effects arise. The first one is the evolution kernel $\mathcal{D}$ which is a part of the  evolution prefactor $\mathcal{R}$, and it is common for all TMDs (TMDPDFs and TMDFFs of various polarizations). The second one is the higher twist corrections to the small-$b$ OPE. These non-perturbative contributions are of essentially different origin and should not be mixed. In particular it is important to realize that the non-perturbative contribution of $\mathcal{D}$ enters Eq.~(\ref{def:pert_TMD}) as a prefactor, while the higher order terms of OPE are added to the convolution integral. Therefore, the structure of non-perturbative corrections to TMD that we keep in mind is the following
\begin{eqnarray}\label{def:initial_TMD_anzatz}
F_{q\ot N}(x,\vec b;\zeta_f,\mu_f)&=&
\exp\Bigg\{\int^{\mu_f}_{\mu}\frac{d\mu'}{\mu'}\gamma\(\mu',\zeta_f\)\Bigg\}\(\frac{\zeta_f}{\zeta_b}\)^{-\mathcal{D}(\mu,\vec b)-\mathcal{D}^{NP}(\vec b)}\times
\\&&\nn\(\sum_j \int_x^1 \frac{dy}{y}\hat C_{q\ot j}\(\frac{x}{y},\vec
b;\mu\)f_{j\ot N}(y,\mu)+f_{q\ot N}^{NP}(x,\vec b;\mu)\).
\end{eqnarray}
Here, $\mathcal{D}^{NP}$ is the non-perturbative addition to the anomalous dimension $\mathcal{D}$, and $f^{NP}$ is the cumulative effect of the higher twist corrections to the small-$b$ OPE. At small (perturbative) $\vec b$, the non-perturbative parts should turn to zero, such that Eq.~(\ref{def:initial_TMD_anzatz}) reproduces Eq.~(\ref{def:pert_TMD}). In the following subsections we construct a minimal non-contradicting anzatz for TMD distributions that respect the study of large-$\beta_0$ approximation.

\subsection{Non-perturbative corrections to the anomalous dimension $\mathcal{D}$}

The non-perturbative  part of the anomalous dimension $\mathcal{D}$ is one of the most studied in the literature and the one for which a general consensus is achieved. 
Usually, the anomalous
 dimension $\mathcal{D}$ is assumed to have  quadratic behavior in the non-perturbative region. 
 As we show in Eq.~(\ref{IRunamb:D}) the quadratic behavior is also suggested by the 
 large-$\beta_0$ approximation.
 A more subtle issue concerns the amount of non-perturbative correction to $\mathcal{D}$, which  can  be very different depending on the  implementation of the TMDs. A check of the renormalon contribution, as provided in this section, gives an estimate of such  correction and it is so useful  for practical implementations.

Let us present the perturbative series for $\mathcal{D}$ in the form 
\begin{eqnarray}
\mathcal{D}(\mu,\vec b)&=&\frac{C_F}{\beta_0} \sum_{n=1}^\infty  (\beta_0 a_s(\mu))^n\( d_n(\mathbf{L}_\mu)+\delta_n(\mathbf{L}_\mu)\),
\label{eq:Dsplit}
\end{eqnarray}
where $d_n\sim n! g^{[n+1]}_0$ can be obtained from Eq.~(\ref{D_AD:n-expression}) and $\delta_n$ is the large-$\beta_0$ suppressed part. The numerical comparison of the large-$\beta_0$ expression Eq.~(\ref{D_AD:summed}) and the exact expression for $\mathcal{D}$ is given in the Tab.~\ref{tab:Dsplit}.
One can see that generally {\it the large-$\beta_0$ expression overestimates the exact numbers}, which is typical for this approximation. 

\begin{table}[t]
\begin{center}
\begin{tabular}{c||l|l|l}
n& $ d_n+\delta_n$ & $d_n$ & $\delta_n$
\\
\hline
1 & $2 \mathbf{L}_\mu$ & $2 \mathbf{L}_\mu$ & 0
\\
2 &$ \mathbf{L}_\mu^2+2.03 \mathbf{L}_\mu^2-1.31 $ & $\mathbf{L}_\mu^2+3.33 \mathbf{L}_\mu+3.11 $ & $-1.30 \mathbf{L}_\mu-4.42$
\\
3 &$\begin{array}{r} 0.67  \mathbf{L}_\mu^3+2.82 \mathbf{L}_\mu^2+0.24 \mathbf{L}_\mu\\-2.41\end{array}$ &
$\begin{array}{r}0.67 \mathbf{L}_\mu^3+3.33 \mathbf{L}_\mu^2+5.56 \mathbf{L}_\mu\\+7.67\end{array}$
&
$-0.51 \mathbf{L}_\mu^2-5.32 \mathbf{L}_\mu-10.$
\end{tabular}
\caption{Numerical comparison of the large-$\beta_0$ component of the  anomalous dimension $\mathcal{D}$ to the exact expression. The coefficients $d_n$ and $\delta_n$ are defined in Eq.~(\ref{eq:Dsplit}).}
\label{tab:Dsplit}
\end{center}
\end{table}

In order  to  study the properties of the large-$\beta_0$ series we introduce a function for its partial sum
\begin{eqnarray}
M_N(\mu,\vec b)=\frac{1}{\beta_0}\sum_{n=1}^N (\beta_0 a_s(\mu))^n d_n(\mathbf{L}_\mu).
\end{eqnarray}
For $N\to \infty$ the sum is divergent, as  discussed in Section \ref{sec:renorm_for_AD}. In order to define $M_\infty$ we consider the Borel transform of $M_N$ as in Sec.~\ref{sec:renorm_for_AD}. To define the Borel integral in Eq.~(\ref{D_AD:summed}), we shift the integration contour, slightly above the real axis. The real part of the integral (i.e. the principal value integral) gives $M_{\infty}$, while the imaginary part represents the errorband for this estimation. The explicit expression for the latter is
\begin{eqnarray}
\delta M(\mu,\vec b)=\frac{2\pi}{\beta_0}\Big[J_0\(\sqrt{\mu^2 \vec b^2}e^{\frac{5}{6}-\frac{1}{2\beta_0a_s(\mu)}}\)-1\Big],
\end{eqnarray}
and the leading behavior at small-$b$ for $\delta M$ is given by the infrared ambiguity Eq.~(\ref{IRunamb:D}).

\begin{table}
\begin{center}
\begin{tabular}{c||l|l|l}
$\mu$=10 GeV  & $b=0.2$ & $b=1.5$& $b=3.0$
\\
\hline
$M_1$ & 0.032 & 0.145 & 0.184
\\
$M_2$ & 0.047 & 0.228 & 0.304
\\
$M_3$ & 0.051 & 0.277 & 0.388
\\
$M_4$ & 0.053 & 0.310 & 0.455
\\
$M_5$ & 0.054 & 0.223 & 0.513 
\\
$M_6$ & 0.054 & 0.354 & 0.567
\\
$M_7$ & 0.055 & 0.372 & 0.622
\\
$M_\infty\pm \delta M$ & $0.055\pm 0.001$  & 
$0.376\pm 0.072$
& $0.577\pm 0.267$
\end{tabular}
\caption{The values of partial sums $M_N$ at several values of $b$. The estimate converge value $M_\infty$ and its error band $\delta M$ are obtained as described in the text\label{tab:Mi}}
\end{center}
\end{table}

\begin{figure}[t]
 \begin{center}
 \includegraphics[width=.45\textwidth,angle=0
]{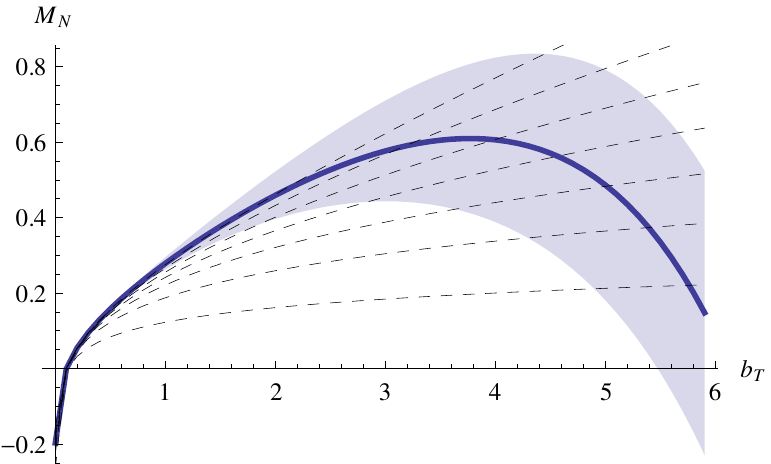}
 \caption{
The dependence of partial sums $M_N$ on $b$ (in GeV$^{-1}$). The dashed lines represent $M_N$ from $N=1$ (bottom line) till $N=7$ (top line). The bold line is the value of $M_\infty$. The shaded area is the error band of $M_\infty$ given by $\delta M$. 
}
\label{fig:convergence}
 \end{center}
 \end{figure}

We investigate the convergence of the partial sums of $M_N$ to its Borel resummed value $M_\infty$, in order to find the scale at which the non-perturbative corrections associated with renormalons become important.
The numerical values of partial sums at $\mu=10$ GeV and at several values of $\vec b$ are presented  in  Tab.~\ref{tab:Mi} . The graphical representation of these values is shown in Fig.~\ref{fig:convergence}.
The convergence of the series is perfect (in the sense that it converges at $M_7$ that is far beyond the scope of modern perturbative calculations) for the range of $b\lesssim 2$GeV$^{-1}$, it becomes weaker at $b\sim 3$ GeV$^{-1}$, and it is completely lost at $b\gtrsim 4$ GeV$^{-1}$. These are the characteristic scales for switching the perturbative and non-perturbative regimes in $\mathcal{D}$. In other words, the perturbative series can be trustful at $b\lesssim 2$GeV$^{-1}$, but completely loses its prediction power for $b\gtrsim 4$ GeV$^{-1}$. The  number $N$ at which convergence is lost depends on the value of $\mu$, however the  interval of convergence in $b$ is $\mu$-independent, e.g. at $\mu=50$ GeV the series converges to $M_8$ in the region $b\lesssim 2$ GeV$^{-1}$, but again loses stability  at $\sim 4$ GeV$^{-1}$.

In order to proceed to  an estimate of the non-perturbative  part of  $\mathcal{D}$ we write it  in the form
\begin{eqnarray}\label{AD_D:complete_form}
\mathcal{D}(\mu,\vec b)=\int_{\mu_0}^\mu \frac{d\mu'}{\mu'}\Gamma_{cusp}(\mu)+\mathcal{D}^{PT}(\mu_0,\vec b)+\mathcal{D}^{NP}(\mu_0,\vec b),
\end{eqnarray}
where $\mathcal{D}^{PT}$ is given by the perturbative expression at $\mu_0$ scale, $\mathcal{D}^{NP}$ encodes the non-perturbative part. The parameter $\mu_0$ depends on $\vec b$ and should be selected such that $a_s(\mu_0)$ is a reasonably small number. The non-perturbative part $\mathcal{D}^{NP}$ is independent on $\mu$ (since the evolution part of $\mathcal{D}$ is renormalon-free) but depends on the choice of $\mu_0$.

In principle, the best value of the  parameter $\mu_0$ can be extracted from the large-$\beta_0$ calculation. Indeed, the resummation of bubble-diagrams modifies the coupling in the interaction vertex, such that a loop integral appears to be naturally regularized in the infrared region. Practically, the effect of such resummation can be presented as a freezing of the coupling constant at large $b$. Particularly popular is the $b^*$ prescription \cite{Collins:1984kg} defined as
\begin{eqnarray}
\mu_0=\mu_b=\frac{C_0}{b^*(\vec b)},\qquad b^*(\vec b)=\frac{\sqrt{\vec b^2}}{\sqrt{1+\vec b^2/b^2_{\text{max}}}},\qquad
C_0=2 e^{-\gamma}.
\end{eqnarray}
At large $b$ the parameter $\mu_0$ approaches $C_0/b_{\text{max}}$, which should be chosen much less then $\Lambda$, i.e. $b_{\text{max}}\ll C_0/\Lambda\sim 4$ GeV$^{-1}$.

For  large-$b$ (say  $b\gtrsim 3$ GeV) the non-perturbative part of $\mathcal{D}$ dominates the perturbative one. The large-$\beta_0$ calculation allows to estimate the leading contribution  (from the side of small-$b$'s) to $\mathcal{D}^{NP}$ from the infrared ambiguity Eq.~(\ref{IRunamb:D}),
\begin{align}
\mathcal{D}^{NP}(\vec b,\m_0)=& c\Lambda^2 \vec b^2 g_D (b,\mu_0),
\end{align}
the  function $g_D$ should be of order of unity at small-$b$ and it depends on the choice of the scale $\mu_0$. Here $\Lambda^2$ is the position of Landau pole and it is expected to be of order  ${\cal O}(\Lambda_{QCD})\sim 250 $ MeV, which implies 
\begin{align}
\label{eq:renDRS}
c\Lambda^2=\frac{\pi C_F e^{5/3}}{2 \b_0} \Lambda^2 \sim 0.075\; {\rm GeV}^2\ .
\end{align}
Since the large-$\beta_0$ approximation overestimates the exact values this number can be considered as an upper bound for non-perturbative input.

In order to estimate the parameters of the $\mathcal{D}$ more accurately, we consider a kind of renormalon subtraction scheme for the anomalous dimension $\mathcal{D}$. We construct a renormalon subtracted expression $\mathcal{D}(\mu,\vec b)=\mathcal{D}^{RS}(\mu,\vec b)$ by explicitly summing the large-$\beta_0$ contribution  in Eq.~(\ref{eq:Dsplit})
\begin{align}\label{AD_D:RS}
\mathcal{D}^{RS}(\mu,\vec b)&=M_\infty(\mu,\vec b)+\frac{C_F}{\beta_0}\sum_{n=1}^\infty (\beta_0 a_s)^n\delta_n(\mathbf{L}_\mu).
\end{align}
The scale $\mu$ here should be chosen such that the logarithm $\mathbf{L}_\mu$ is reasonably small, otherwise the large-$\beta_0$ expansion is significantly violated. Using the model Eq.~(\ref{AD_D:RS}) we fit the parameters of Eq.~(\ref{AD_D:complete_form}) at $\mu=10$ GeV in the range $b<3$ GeV, with $g_D=$ constant$\equiv g_K$, at all known perturbative orders.
It appears that the result is very stable with respect to $b_{\max}$ whose best value we find to be
\begin{eqnarray}
b_{\max}\simeq (1.2\pm 0.1) \text{~GeV}^{-1}.
\end{eqnarray}
Concerning the  non-perturbative part, it appears to be lower then the crude estimation Eq.~(\ref{eq:renDRS}) and actually consistent with 0,
\begin{eqnarray}
g_K\simeq (0.01\pm 0.03) \text{~GeV}^2.
\end{eqnarray}
This value is generally smaller  then the typical values presented in the literature, e.g. Ref.~\cite{Aybat:2011zv} quotes $g_K\simeq0.17$GeV$^2$, Ref.~\cite{Echevarria:2014xaa} quotes $g_K\simeq 0.045\pm 0.005$GeV$^2$. But Ref.~\cite{D'Alesio:2014vja} finds $g_K$ consistent with 0, which agrees with the present findings. However, one should take into account that contrary to standard fits, the present considerations are purely theoretical. Moreover in  fits with experimental data, one should consider the extra non-perturbative part of the TMD distribution itself (which is discussed in the next section).

Finally, we comment on the possibility of a more sophisticated  renormalon subtraction  scheme as in the MSR scheme  of \cite{Hoang:2009yr}. In this scheme, one provides a  subtraction of the renormalon from a perturbative series  which depend on an additional scale $\mu_R$. The new renormalon subtraction scale can result into large logarithms which, in turn, should be resummed. Such a consideration can result in more accurate restrictions on parameters.

\subsection{Renormalon consistent ansatz for TMDs}
\label{sec:MODEL}

The non-perturbative corrections to the matching coefficients are necessary for all analysis which include low energy data. These corrections have not been  deeply studied in QCD theory and up to now, only a phenomenological treatment has been provided. In this section, we present a consistent ansatz that interpolates the perturbative small-$b$ part of a TMD distribution with an entirely Gaussian exponent at large-$b$. The presented ansatz takes into account the lessons learned from the study of renormalon singularities and formulated in Sec.~\ref{subsec:lessons}. 

The renormalon contribution accounts the leading power correction (see detailed explanation e.g. in \cite{Dokshitzer:1995qm,Beneke:2000kc,Braun:2004bu}). Thus, the small-$b$ expansion of the TMD distribution, that includes this power correction, has a form
\begin{align}
\hat F_{q\ot N}(x,\vec b;\mu)=\sum_j\int_x^1 \frac{dy}{y}\(\hat C_{q\ot j}(y,\vec b;\mu)+y g_q^{in} \vec b^2  C^{ren}_{q\ot j}(y,\vec b)\) f_{j\ot N}\(\frac{x}{y},\mu\)+{\cal O}(\vec b^4),\label{eq:CansatzF}
 \\
\hat D_{q\to N}(z,\vec b;\mu)=\sum_j\int_z^1 \frac{dy}{y}\(\hat{\mathbb{C}}_{q\to j}(y,\vec b;\mu)+\frac{g_q^{out} \vec b^2}{y}C^{ren}_{q\to j}(y,\vec b)\) d_{j\to N}\(\frac{z}{y},\mu\)+{\cal O}(\vec b^4),
\label{eq:CansatzD}
\end{align}
where the LO coefficient function of the renormalon contribution was calculated in Sec.~\ref{sec:matching_at_b0} and reads
\begin{eqnarray}
C^{ren}_{q\ot q}(x,\vec b)=C^{ren}_{q\to q}(x,\vec b)=2\bar x+\frac{2x}{(1-x)_+}-\delta(\bar x)\(\mathbf{L}_\Lambda+\frac{2}{3}\).
\end{eqnarray}
The constants $g_q^{in,out}$ are of order $c\Lambda^2$ within the large-$\b_0$ approximation, however the actual value should be estimated from data. The non-perturbative scale $\Lambda$ is the same as in the case of the evolution kernel. The contribution presented here  is at LO, and as such has not $\mu$-dependence. The $\mu$-dependence of higher perturbative orders can in principle be calculated, using  the evolution equation for TMD and the related integrated distribution.

At larger values of $b$ Eq.~(\ref{eq:CansatzD}) is corrected by the higher orders of the OPE, and at a particular scale $B$ (which defines the convergence radius of small-$b$ OPE Eq.~(\ref{coeff_powers})) it is replaced by a single and entirely non-perturbative function. It is commonly assumed that at large-$b$ the TMD distribution has Gaussian behavior. This is also supported by the phenomenological studies of low-energy data (see e.g. Ref.~\cite{Schweitzer:2010tt} for a study dedicated to this issue). The interpolation of a Gaussian with the small-$b$ matching Eq.~(\ref{eq:CansatzF}-\ref{eq:CansatzD}) should take into account the previously formulated  demands on the power corrections. In particular, we have the following two guidelines:
\begin{description}
\item(i) In order to be consistent with the general structure of OPE, the interpolation should be done under the convolution integral.
\item(ii) According to the structure of renomalon singularities, the powers of $\vec b^2$ should be always supplemented by $x$ (for PDF) and $z^{-1}$ (for FF).
\end{description}
A viable model, which takes into account both these points, can have the form 
\begin{align}\label{eq:Cansatzv2}
&\hat F_{q\ot N}(x,\vec b;\mu)=
\\\nn&\sum_j \int_x^1\frac{dy}{y}e^{-g_b y\vec b^2}\(\hat C_{q\ot j}(y,\vec b;\mu)+y g_q \vec b^2  \(C^{ren}_{q\ot j}(y,\vec b)+\delta(\bar y)\frac{g_b}{g_q}\)\)
f_{j\ot N}\(\frac{x}{y},\mu\),
\\\label{eq:CansatzFv2}
&\hat D_{q\to N}(z,\vec b;\mu)=\\
\nn
&\sum_j\int_{z}^1 \frac {dy}{y}e^{-g_b \vec b^2/y}\(\hat{\mathbb{C}}_{q\to j}(y,\vec b;\mu)+\frac{g_q \vec b^2}{y}\(C^{ren}_{q\to j}(y,\vec b)+\delta(\bar y)\frac{g_b}{g_q}\)\) d_{j\to N}\(\frac{z}{y},\mu\).
\end{align}
The
inclusion of the perturbative and power corrections modifies the Gaussian shape  differently for PDF and FF kinematics.

\begin{figure}[t]
 \begin{center}
 \includegraphics[width=.45\textwidth, angle=0
]{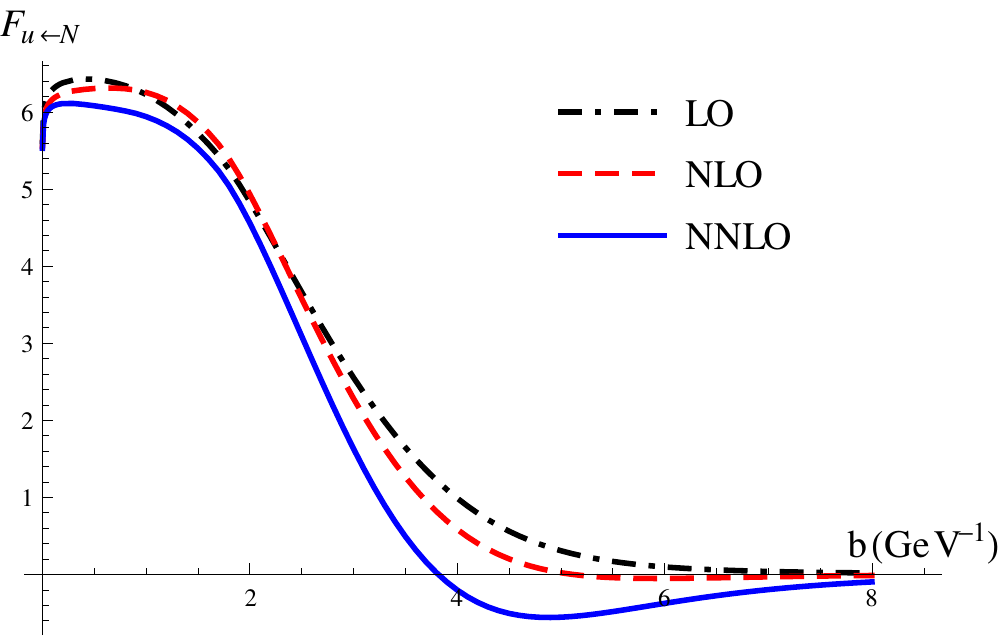}
 \includegraphics[width=.45\textwidth, angle=0
]{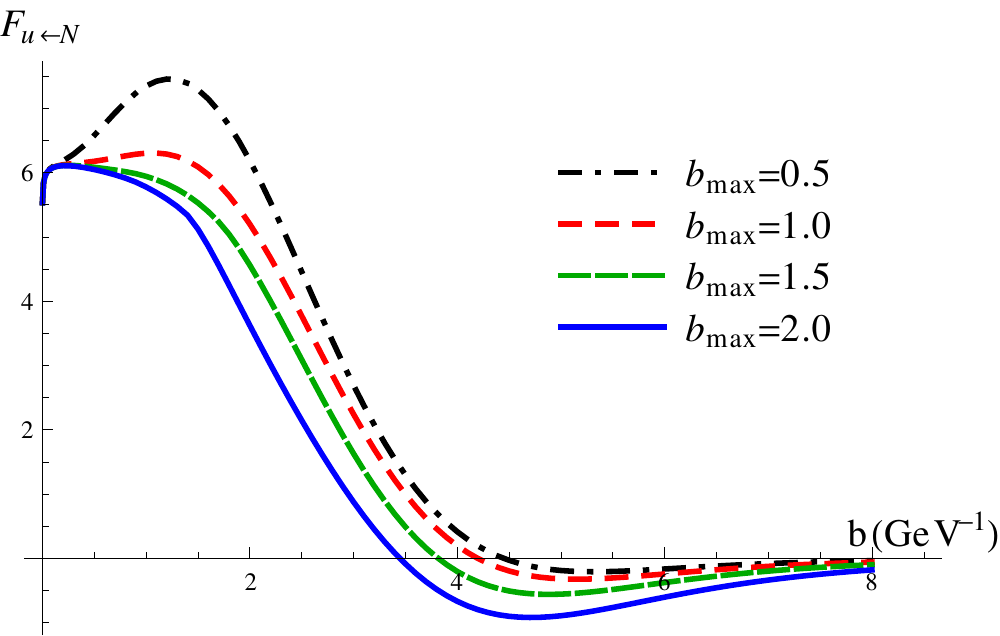}
 \caption{The TMDPDF $ \hat F_{u\ot p}(x,\mathbf{L}_\mu)$  as in the model of Eq.~(\ref{eq:Cansatzv2})  (the up-quark PDF is taken from MSTW \cite{Martin:2009iq,Harland-Lang:2014zoa}, at $x=0.1$,  $\Lambda=0.25$ GeV,  $N_f=3$, $g_b=.2$ GeV$^{-2}$, $g_q=0.01$ GeV$^{-2}$, $\mu=C_0/b^*$)  as a function of the parameter $b$ in GeV$^{-1}$ units. On the left panel we show consequently curves for LO, NLO and NNLO matching coefficients ($b_{max}=1.5$ GeV$^{-1}$ is used). On the right panel we present NNLO curve at several values of 
$b_{max}$ in units of GeV$^{-1}$.}
\label{fig:Chat1}
 \end{center}
 \end{figure}

\begin{figure}[t]
 \begin{center}
 \includegraphics[width=.45\textwidth, angle=0
]{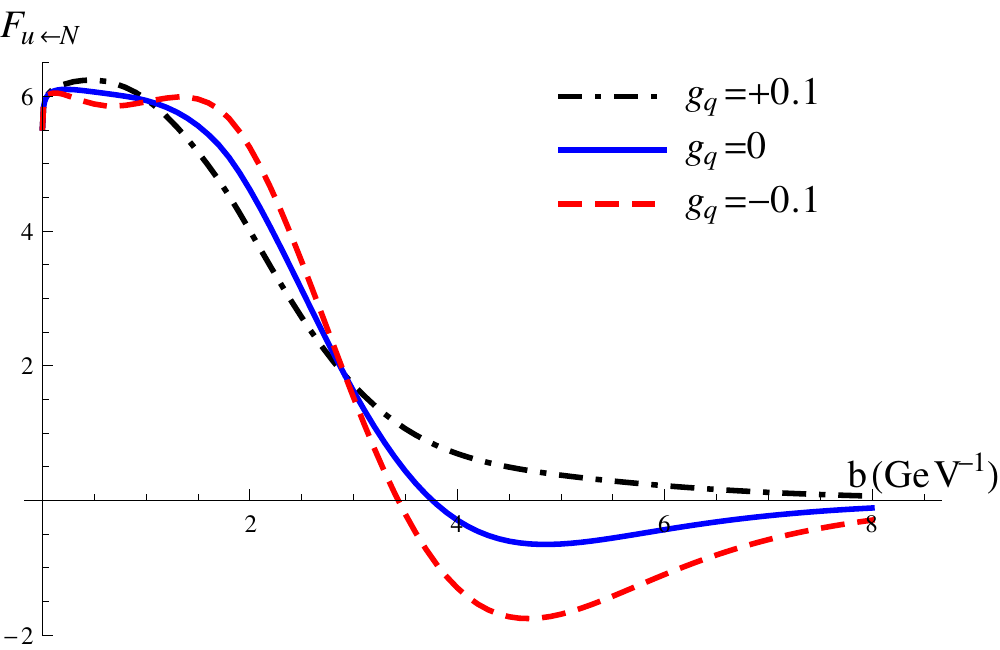}
  \includegraphics[width=.45\textwidth, angle=0
]{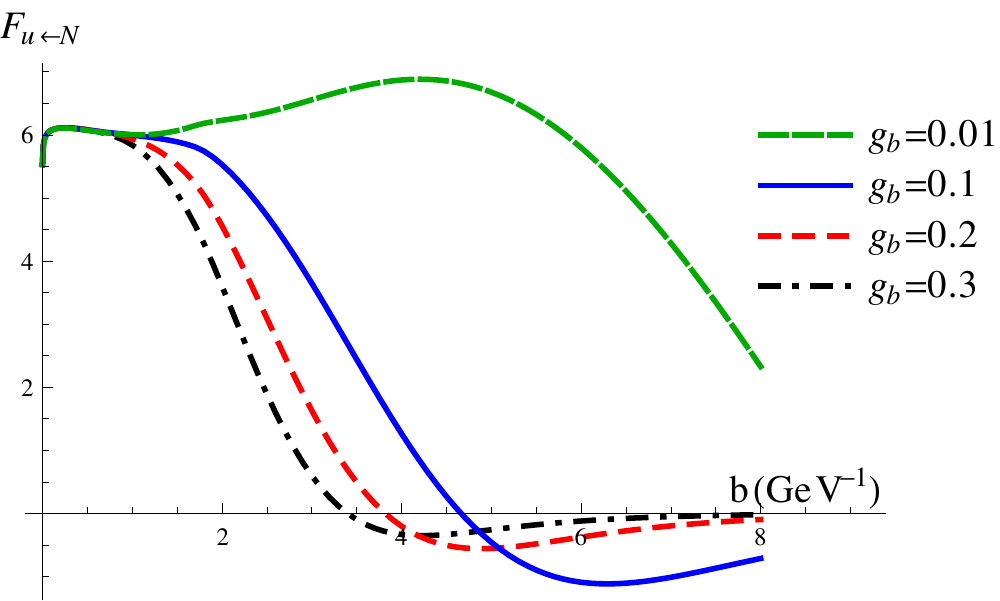}
 \caption{The  TMDPDF $ \hat F_{u\ot p}(x,\mathbf{L}_\mu)$  as in the model of Eq.~(\ref{eq:Cansatzv2}) at NNLO (PDF from MSTW \cite{Martin:2009iq,Harland-Lang:2014zoa}, and with $x=0.1$,  $\Lambda=0.25$ GeV,  $N_f=3$, $\mu=C_0/b^*$ with $b_{\max}=1.5$ GeV$^{-1}$) as a function of the impact parameter $b$ in GeV$^{-1}$ units.  
On the left panel we show several possible choices of $g_q$ in GeV$^{2}$ at fixed $g_b=.2$GeV$^2$. 
On the right panel we show several possible choices of $g_b$ in GeV$^{2}$ at fixed $g_q=.01$GeV$^2$. All curves are at NNLO.
\label{fig:ChatgQ}}
 \end{center}
 \end{figure}

\begin{figure}[t]
 \begin{center}
 \includegraphics[width=.5\textwidth, angle=0
]{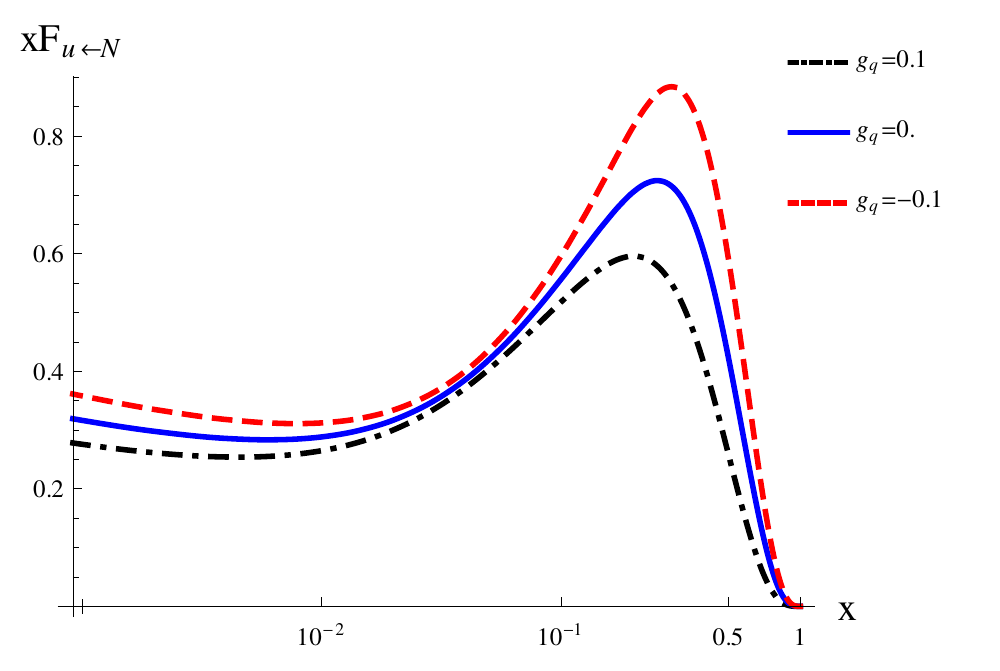}
 \caption{The function $ x \hat F_{u\ot p}(x,\mathbf{L}_\mu)$  as in the model of Eq.~(\ref{eq:Cansatzv2}) at NNLO (PDF from MSTW \cite{Martin:2009iq,Harland-Lang:2014zoa}, as a function of $x$. The other inputs are fixed as $\Lambda=0.25$ GeV, $N_f=3$,  $\mu=C_0/b^*$, $b=b_{max}=1.5$ GeV$^{-1}$ and $g_b=0.2$GeV$^2$. We show the curves at different values of $g_q$.
\label{fig:ChatgX3}}
 \end{center}
 \end{figure}

In the figures \ref{fig:Chat1}-\ref{fig:ChatgX3}  we illustrate several features of the renormalon consistent ansatz that we propose. In all the plots we fix the $\mu$ scale at the value $\mu=\mu^*=C_0/b^*$. In Fig.~\ref{fig:Chat1}-left we show that the change of $\hat F$ with respect to the perturbative order of matching coefficient.  On the right hand side of Fig.~\ref{fig:Chat1} we show the dependence on the choice of the scale $b_{max}$, which we find very mild for $1$~GeV$^{-1}\lesssim b_{max}\lesssim 2$ GeV$^{-1}$.

The shape  of the TMDs can strongly depend of the values of  the non-perturbative constants $g_{b,q}$  for $b\geq 2$ GeV$^{-1}$  as shown in Fig.~\ref{fig:ChatgQ}-\ref{fig:ChatgX3}. The values  used in plots parameters are inspired by the fit in~\cite{D'Alesio:2014vja}. However, they can also change in a real fit with the present model. For $b\leq 1$ GeV$^{-1}$  the non-perturbative model does not really affect the $x-$behavior of the TMD. In  Fig.~\ref{fig:ChatgX3} we  show instead that for instance at $b\sim 1.5$  GeV$^{-1}$ the model parameter can start to have their impact.

\begin{figure}[t]
 \begin{center}
 \includegraphics[width=.45\textwidth, angle=0
]{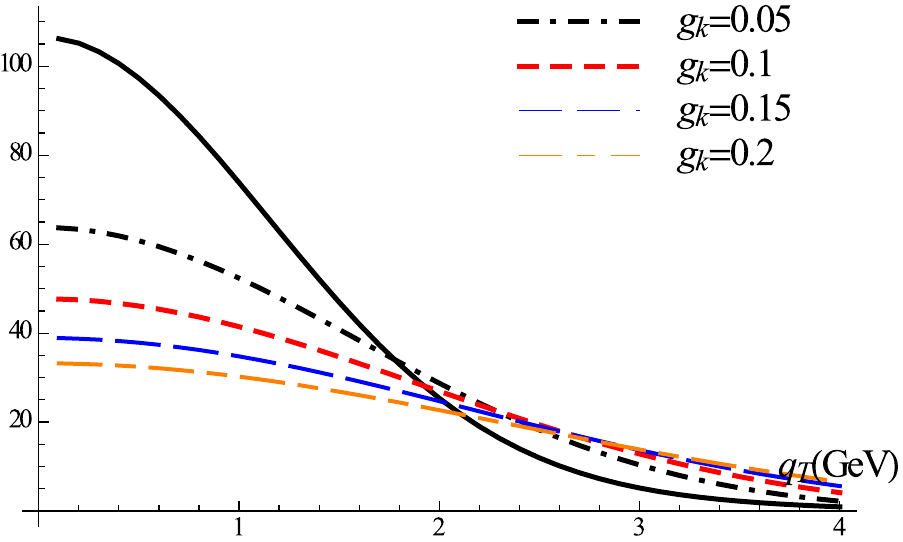}
 \includegraphics[width=.45\textwidth, angle=0
]{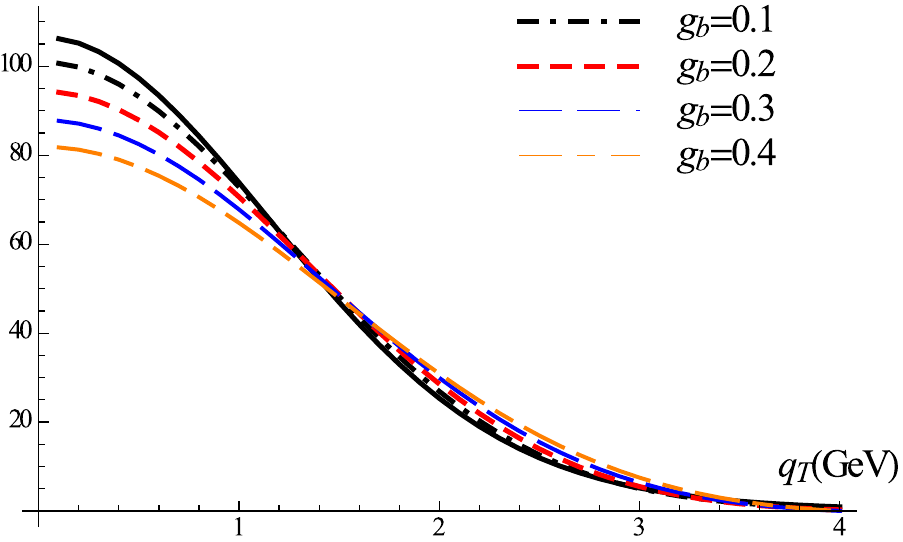}
 \includegraphics[width=.45\textwidth, angle=0
]{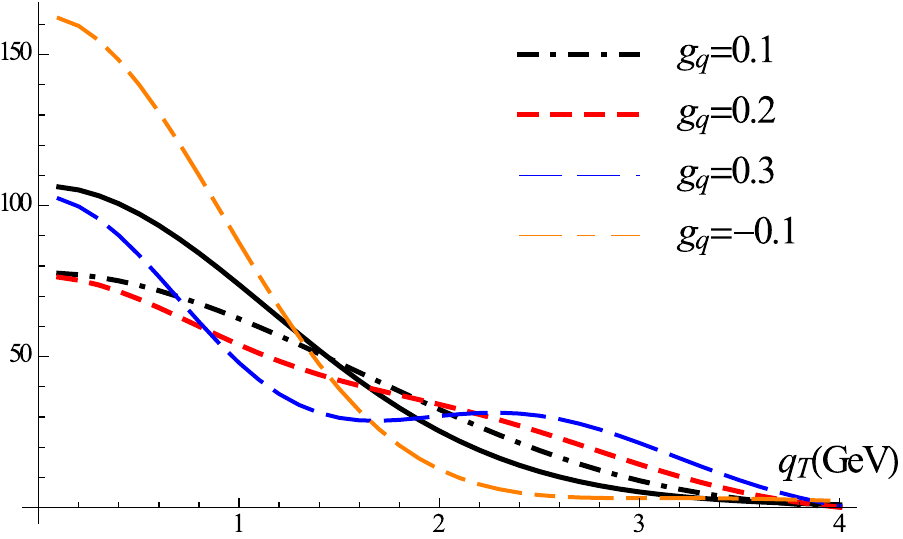}
 \caption{The plots of Drell-Yan cross-section $p+p\to \gamma+X$ $d\sigma/dQ^2dy dq_T^2$ at $\sqrt{s}=100$GeV, $Q=10$GeV and $y=0$, evaluated using the renormalon ansatz.  The impact of different parameters is demonstrated. The black line is the reference curve with all parameters set to 0.  The other inputs are fixed as $\Lambda=0.25$ GeV, $N_f=3$,  $\mu_0=C_0/b^*$, $b_{max}=1.5$ GeV$^{-1}$. All curves are at NNLO.
 \label{fig:Xsec1}}
 \end{center}
 \end{figure}

\begin{figure}[t]
 \begin{center}
 \includegraphics[width=.45\textwidth, angle=0
]{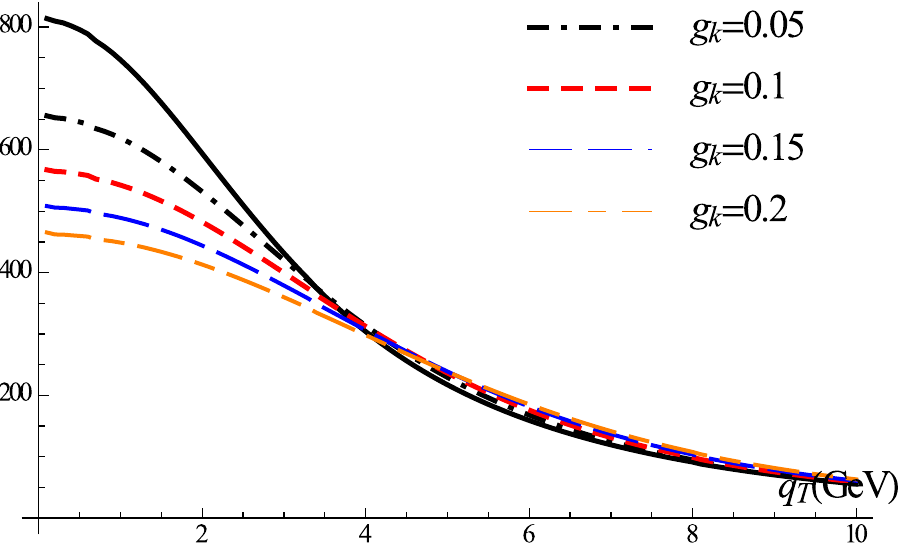}
 \includegraphics[width=.45\textwidth, angle=0
]{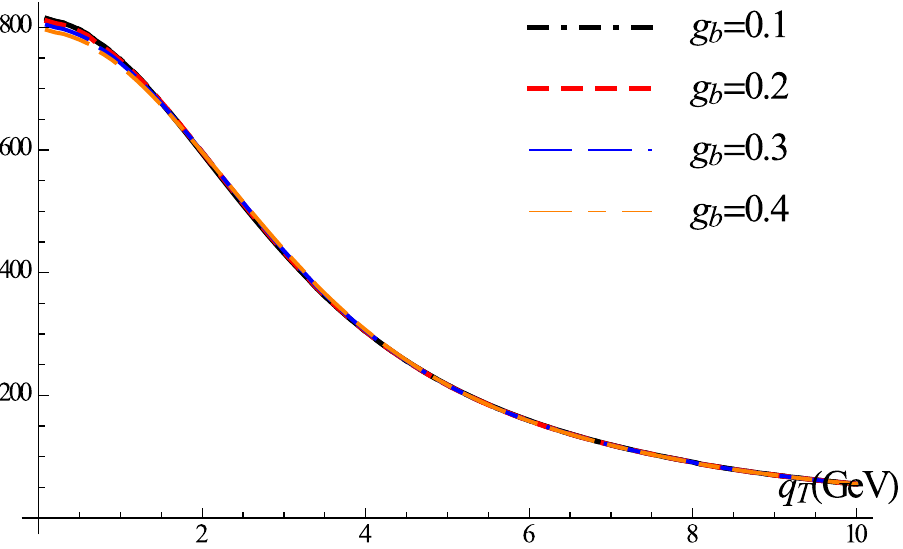}
 \includegraphics[width=.45\textwidth, angle=0
]{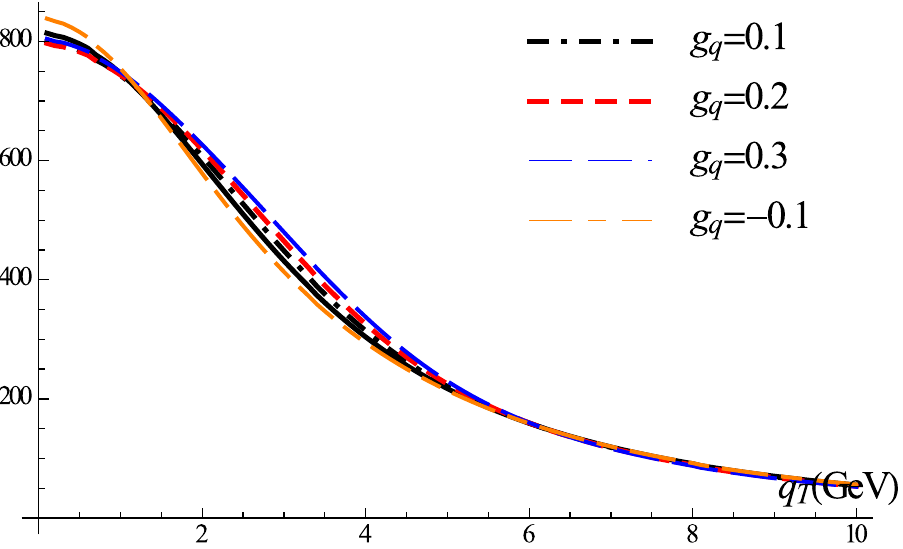}
 \caption{The plots of Z-boson production cross-section $p+p\to \gamma+X$ $d\sigma/dQ^2dy dq_T^2$ at $\sqrt{s}=1.96$TeV, $Q=M_Z=91.18$GeV and $y=0$, evaluated using the renormalon ansatz. The impact of different parameters is demonstrated. The black line is the reference curve with all parameters set to 0.  The other inputs are fixed as $\Lambda=0.25$ GeV, $N_f=3$,  $\mu_0=C_0/b^*$, $b_{max}=1.5$ GeV$^{-1}$. All curves are at NNLO.
\label{fig:Xsec2}}
 \end{center}
 \end{figure}

The cross-section built from TMDs in the form (\ref{eq:Cansatzv2}-\ref{eq:CansatzFv2}) and the evolution kernel (\ref{AD_D:complete_form}) is dependent on the parameters $g_K$, $g_b$ and $g_q$. While, the parameter $g_K$ is strongly universal, the parameters $g_b$ and $g_q$ are separate for TMDPDFs and TMDFFs, as well as, different for different flavors. Within the cross-section the dependence on these parameters is smoothed to a more-or-less similar shape (especially for parameters $g_b$ and $g_K$). However, the dependence on these parameters is clearly distinguishable at different energies. As an example, we show the  Drell-Yan cross-section in Fig.~\ref{fig:Xsec1} and the Z-boson cross section in Fig.~\ref{fig:Xsec2} with some typical values of the experimental energies.
 While the corrections to the Z-boson production are dominated  by $g_K$, at low energies all parameters can  compete.  In actual experiments the Z-boson production  is only minimally affected by non-perturbative effects, so  in actual fits it may happen that the value of $g_K$ is compatible with zero, while the other parameters provide the  expected minimal correction (this is for instance the case of the fit in Ref.~\cite{D'Alesio:2014vja}). This yields that an estimate of the nature of the TMDs non-perturbative part cannot be done  just using the Z-boson production,  but needs also data from low energy physics.
  We postpone to a future work a   comparison with data of the  model that we have presented here.

To conclude this section, we observe  that in the literature we have not found any non-perturbative input for TMDs fully consistent with the demands dictated by the power analysis presented here. For instance  the $b^*$-prescription which is used in many  phenomenological analysis \cite{Aybat:2011zv,Collins:2014jpa,RESBOS} is inconsistent with Eq.~(\ref{eq:CansatzF}). Within the $b^*$-prescription the higher-twist corrections are simulated by replacing $b\to b^*$, and including an additional non-perturbative factor as
\begin{eqnarray}\label{eq:b*_model}
\hat F^{\text{$b^*$-presc.}}_{q\ot N}(x,\vec b;\mu)=\sum_j  e^{g_{j/N}(x,\vec b)}\int_x^1\frac{dy}{y}\hat C_{q\ot j}(y,b^*;\mu)f_{j\ot N}\(\frac{x}{y},\mu\),
\end{eqnarray}
and similarly for TMDFF. This expression violates both guidelines formulated before Eq.~(\ref{eq:Cansatzv2}). Considering the small-$b$ expansion of $\hat C(b^*)$ in Eq.~(\ref{eq:b*_model}),
\begin{eqnarray}\label{eq:b*_model_expanded}
&&\hat F^{\text{$b^*$-presc.}}_{q\ot N}(x,\vec b;\mu)|_{{\rm small-}b}\simeq
\sum_j  \int_x^1\frac{dy}{y}\Big[\hat C_{q\ot j}(y,\vec b;\mu)
\\&&\quad+\nn\frac{a_s(\mu)C_F \vec b^2}{b_{\max}^2}\(\frac{2y}{(1-y)_+}+\bar y-\delta(\bar y)\(\mathbf{L}_\mu-\frac{3}{2}\)\)+\delta(\bar y)\vec b^2 g''_{j/N}(x,0)\Big]f_{j\ot N}\(\frac{x}{y},\mu\),
\end{eqnarray}
one does not reproduce  Eq.~(\ref{eq:CansatzF}). The main difference comes from the general power scaling, $x\vec b^2$ vs. $\vec b^2$, see point (ii). The point (i) is violated by the non-perturbative exponent that is generally $x$-dependent and positioned outside of convolution integral (although, we should appreciate that in most application it is taken $x$-independent).

\section{Conclusion}

In this work, we have studied the non-perturbative properties associated with renormalons for the soft function and unintegrated matrix elements. With this aim, we have evaluated all constituents of TMD distributions (soft factor, matching coefficient and anomalous dimensions) within the large-$\beta_0$ approximation. The (factorial) divergences of the large-$\beta_0$ series are associated with the renormalon contribution and allow to estimate the leading non-perturbative contributions. We have found two independent renormalon structures in the perturbative description of TMD: the soft function and small-$b$ matching coefficients.

The consideration of the soft function allows to fix the power behavior of the evolution kernel of TMDs. We show the evidence of infrared renormalons at $u=1,2,..$ ($u$ being the Borel parameter). 
Our results agree with the analysis of the power corrections to factorized cross-section made in \cite{Becher:2013iya}. It also supports the popular assumption about a quadratic power correction to the TMD evolution kernel. However, the impact of the non-perturbative corrections is estimated to be not very significative for experiments where TMDs are evaluated at scales higher than a few GeV.

The nature of the renormalon contribution to the evolution kernel is peculiar, in the sense that it is generated by the non-perturbative part of  a matrix element. In some aspects, this is very similar to the renormalon contribution to heavy quark masses. We have discussed also an ansatz which implements a consistent renormalon subtraction for the TMD evolution kernel,  which  can be useful for phenomenology.

The most promising conclusion of the paper comes from the analysis of  the renormalon contribution to the small-$b$ expansion of TMDs. The discussion of these results  can be found in Sec.~\ref{subsec:lessons}. We demonstrate that the power corrections to small-$b$ behave as a function of $x\vec b^2$ for TMDPDFs and as $\vec b^2/z$ for TMDFFs. This observation should have a significant impact on the joined TMDPDF -- TMDFF phenomenology. Additionally, the  large-$\beta_0$ computation  unveils  the form of $x-$dependence for the leading power correction to the small-$b$ matching. This behavior should be incorporated in realistic and consistent models for TMDs.

We have discussed and formulated the demands on a phenomenological ansatz to incorporate all collected information. We find that typical models for the non-perturbative part of TMDs, discussed in the literature, are inconsistent with our conclusions, mainly, due to the naive assumption that the combined powers corrections are largely functions of $\vec b^2$ (contrary to $x\vec b^2$). In eqns.(\ref{eq:Cansatzv2}-\ref{eq:CansatzFv2}) we construct a simple ansatz that interpolates the Gaussian low-energy model for TMDs with the perturbative small-$b$ regime accounting formulated demands.  We postpone to a future work the fit of available data using the presented results.

\section*{Acknowledgements}
We thank Vladimir Braun for numerous discussions and useful comments. We thank  the Erwin Schrödinger International Institute for Mathematics and Physics (ESI, Vienna) for kind hospitality during the summer 2016 and for propitiating nice discussions on this work.
I.S. is supported by the Spanish MECD grant FPA2014-53375-C2-2-P and FPA2016-75654-C2-2-P.

\end{document}